\documentclass{elsart}
\usepackage{amsmath}
\usepackage{graphics}

\DeclareMathOperator{\circul}{circ}
\DeclareMathOperator{\mmod}{mod}

\newcommand{\ben}{\begin{equation}}
\newcommand{\een}{\end{equation}}
\newcommand{\be}{\begin{equation*}}
\newcommand{\ee}{\end{equation*}}

\begin{document}

\runauthor{Voorhees and Beauchemin}

\begin{frontmatter}
\title{Point Mutations and Transitions Between Cellular Automata Attractor Basins}

\author[Athabasca]{Burton Voorhees\thanksref{contact}}
\author[UofA]{Catherine Beauchemin}

\thanks[contact]{Author to whom correspondence should be addressed. Phone: (780)482-3597, E-mail: \texttt{burt@athabascau.ca}}

\address[Athabasca]{Center for Science, Athabasca University, 1 University Drive, Athabasca, AB, CANADA, T9S 3A3}
\address[UofA]{Department of Physics, University of Alberta}

\begin{abstract}
We consider transformations between attractor basins of binary cylindrical cellular automata resulting from mutations.  A $\tau$-point mutation of a state consists in toggling $\tau$ sites in that state.  Results of such mutations are described by a rule-dependent probability matrix.  The structure of this matrix is studied in relation to the structure of the state transition diagram and several theorems relating these are proved for the case of additive rules.  It is shown that the steady state of the Markov process defined by the probability matrix is always the uniform distribution over the state transition diagram.  Some results on eigenvalues are also obtained.
\end{abstract}

\end{frontmatter}

\section{Introduction}
In one form or another, the dialectical view of change formalized by Hegel in the early nineteenth century has dominated most of nineteenth and twentieth century thought. In line with this view, evolutionary processes were initially conceptualized in terms of gradual optimization in which small variations, or mutational changes struggled for survival in environments with limited resources. In recent theorizing about complex adaptive systems, however, the idea of sudden mutational change has come to play a significant role.  This is change that can occur suddenly, in apparently unpredictable jumps, rather than as a gradual transformation of quantity into quality.

Paradigmatic examples of such mutational change are state transitions in quantum systems and genetic mutations, and mathematical models exhibiting the potential for such mutational jumps have proliferated.  In continuous dynamical systems, research has demonstrated there may be many metastable attractors with random jumps between attractors induced by noise \cite{ref3,ref4,ref1,ref2} and, under certain circumstances, these jumps may be controllable \cite{ref5}.  Similar results on noise induced transitions between metastable states have been obtained by Antonelli and Zatawniak \cite{ref6}.  In a model of the evolution of a dimorphic clone in the presence of both internal developmental noise and environmental fluctuations they show that stationary solutions ``are segregated into disjoint invariant sets, providing clonal type stability, growth canalization, and variability within each clonal type... [while] the interaction between the environmental and developmental noise can trigger transitions... from one clonal type to another.''

Genetic mutation and evolution have been taken as a generic model in many studies of change in complex systems. Work at the forefront of complex systems research has focused on such mutational jumps \cite{ref8,ref11,ref12,ref7,ref10,ref9}, emphasizing a new approach to evolutionary change that Crutchfield \cite{ref8} has called epochal evolution.  An important aspect in theorizing about epochal evolution processes is categorization of abstract ``genotypes'' into fitness classes, with all genotypes in a given class having more or less equal fitness.  Populations are described by a probability distribution over these classes, with selection acting between, but not within, each class.

Since the space of possible genotypes is extremely large, what Scott \cite{ref13} terms ``immense'', at any given historical period only a small number of genotypes will actually be manifest in a population.  Innovations arise when random drift within an existing fitness class ``discovers'' a portal to a previously unoccupied class with higher fitness.  Exploitation of the advantage of this new class of genotypes leads to rapid change in the population distribution with the highest fitness class dominating --- a new evolutionary epoch has arisen.

An explicit feature of the theory of epochal evolution is the idea of modularity in the space of genotypes (or, more generally, in an appropriate state space).  The idea of modularity has been present in ecology at least since May's seminal work on stability and complexity of ecosystems \cite{ref14}.  It was May's suggestion that in the case of complex adaptive systems there will be only very small regions in the system parameter space where the system has long term stability.  In genetic terms, it might be posited that only certain prototypic genotypes are compatible with an organism's survival and that fitness classes can be defined as the classes of genotypes that are related to these prototypes by neutral mutations. (It must be emphasized, of course, that the idea of a prototype for a fitness class is an idealization.  There may well be no genotype in a class that actually matches the prototype, which can be taken as a fictional genotype optimized to an idealized environment in which there are no fluctuations of environmental parameters.)

Portals between fitness classes appear at points where a jump between one fitness class and another is possible via only a single or a small number of mutations.  While it is usually assumed that each genotype within a given fitness class has exactly the same fitness as every other genotype in the class, this is not a necessary assumption. All that is required is that the time average fitness of each genotype in any given class, when weighted by the spectrum of environmental fluctuations, be equal.

For example, if the environment can be modelled as fluctuating between $n$ different states labelled $e_j$, $1 \le j \le n$, and a given fitness class $F(r)$ contains $m$ genotypes $g_i$, $1 \le i \le m$, then the $m\times n$ matrix $F_{ij}(r)$ with $i$,$j$ entry equal to the fitness of genotype $g_i$ in environment $e_j$ describes the overall fitness of the class $F(r)$.  The condition that $F(r)$ be a fitness class is that
\ben
\label{eq1.1}
\sum\limits_{j=1}^{n} F_{ij}(r) p_j = \sum\limits_{j=1}^{n} F_{kj}(r) p_j \ ,
\een
for all $i$ and $k$, where $p_j$ is the probability that environmental conditions $e_j$ will occur.  Then consideration can focus on the dynamics introduced by mutations that result in transitions between fitness classes.

One means of representing a system with modular classes undergoing epochal evolution is graphically, as a set of vertices labelled either by genotypes or by fitness classes, with an edge connecting two vertices if there is a mutation that relates the corresponding genotypes or fitness classes.  This representation allows application of the tools of graph theory and network dynamics \cite{ref17,ref18,ref15,ref16,ref19}.  Network dynamics, in both discrete and continuous forms, is emerging as a major mathematical tool in many areas of biology, social science, and economics.  In real cases, however, the networks encountered tend to be highly complex and analytically intractable.  Thus, the study of simple ``toy'' models has become important as a means of gaining insight into real world cases where general underlying principles and laws might be masked by the high level of complexity.  Two cases of simple model systems are found in cellular automata (CA), and in random Boolean networks \cite{ref21,ref7,ref20}.  This paper considers cellular automata as potential models of complex systems with modular network structures that are subject to mutational transitions.

As a basis for modelling mutational jumps between fitness classes (or more generally, between modular elements arising in network dynamics), however, cellular automata are insufficient.  While the attractor basins of a cellular automaton can be used as models of modular units such as fitness classes, there is no mechanism providing for transitions between basins.  Cellular automata dynamics are completely deterministic --- every state lies in a basin of attraction, and iteration of an update rule takes each state to a specific attractor.  Thus, interest in cellular automata has usually focused on elucidation of attractor basins \cite{ref23,ref22}, exploration of space-time patterns generated \cite{ref25,ref24}, and determination of the mathematical properties of various types of rules \cite{ref26,ref27,ref28} (see \cite{ref21} for extensive review).

Recently, however, Wuensche \cite{ref29} has pointed out the possibility of using cellular automata attractor basins as models of modular sub-networks and studying either changes in the network topology of state transition diagrams introduced by perturbations of the generating rules, or transitions between attractor basins induced by mutations in the states themselves.  The first approach has been explored using probabilistic cellular automata \cite{ref30}, but little attention has focused on the second.

In the present paper this second approach is taken up with a study of mutationally induced transitions between basins of attraction of simple cellular automata.  There are two ways to view such transitions.  The first, which is the focus of this paper, is to introduce point mutations by toggling one of more sites in a given state and study the nature of the transitions this introduces between attractor basins of a given CA rule.  The second method, to be treated in a subsequent paper, is based on the specification of a probability matrix that arbitrarily fixes the probability of a transition between distinct basins of attraction. In this approach, one studies the effect of such transitions and their relation to a defined cost, or fitness function.

Consideration is limited to binary valued ``cylindrical'' cellular automata \cite{ref31}, that is, cellular automata rules defined on binary strings of fixed length with periodic boundary conditions.  In contrast to the usual convention, however, in which neighbourhoods are taken as symmetric about a central mapping site, left-justified neighbourhoods are used in this paper.  That is, if $\{i,i+1,\dots,i+k-1\}$ denotes a $k$-site neighbourhood, then the value assigned by this neighbourhood at the next iteration of the CA rule appears at site $i$.  This has certain advantages when considering mappings of half-infinite binary strings as maps of the unit interval.  The main effect of this different neighbourhood choice is that some cycle periods are changed from the symmetric neighbourhood case.

It is also assumed that mutations occur on a much faster time scale than CA rule iterations.  This means that the entire attractor basin is important rather than just the attractor itself.  Physically this corresponds to systems whose natural dynamics operate on a time scale that is orders of magnitude slower than environmental fluctuations that might induce mutations.  The opposite case, in which the cellular automata dynamics operates on a much faster time scale than that of mutation leads to a situation in which only the attractors are relevant since any mutation to a state not on an attractor will iterate quickly to the attractor.

\section{Point Mutations on $E_n$}
\label{sec2}
Given an $n$-digit binary string $\mu$ a mutation $\mu\rightarrow\mu'$ is produced by randomly toggling one or more of the digits of $\mu$.  A $\tau$-point mutation corresponds to toggling $\tau$ digits.  The effect of such a mutation on elements of $E_n$, the state space of $n$-digit binary sequences, is described by a $2^n\times2^n$ (0,1) matrix $T_n(I,\tau)$ defined by
\ben
\label{eq2.1}
\left[T_n(I,\tau)\right]_{ij} =
\begin{cases}
1 & \text{$\exists$ a $\tau$-digit toggle $j \rightarrow i$} \\
0 & \text{otherwise}
\end{cases} \ ,
\een
where the indices $i$ and $j$ are expressed as $n$-digit binary strings.  If $j_0 \dots j_{n-1}$ is the binary form of $j$ then toggling $\tau$ digits is equivalent to the site-wise addition $\mmod(2)$ of an $n$-digit string containing $\tau$ ones.  It is easy to see that $T_n(I,0)$ is the $2^n \times 2^n$ identity matrix $I$ while $T_n(I,n)$ is the $2^n \times 2^n$ anti-identity $I_n^*$.  $T_n(I,\tau)$ is symmetric and each row and each column of $T_n(I,\tau)$ contains
\be
\binom{n}{\tau} = \frac{n!}{\tau!(n-\tau)!}
\ee
ones.

Graph theoretically, $T_n(I,\tau)$ is the adjacency matrix of the $n$-hypercube $H_n(\tau)$ with edges connecting vertices that are separated by Hamming distance $\tau$.  Note also that since $T_n^2(I,1)$ allows for the case in which the same site is toggled twice, $T_n(I,2) \ne T_n^2(I,1)$.  Instead, $2T_n(I,2) = T_n^2(I,1) - nI$.  
Since each row and column of $T_n(I,\tau)$ contains the same number of ones, the matrix
\ben
\label{eq2.2}
\overline{T}_n(I,\tau) = \frac{1}{\binom{n}{\tau}}T_n(I,\tau)
\een
is a probability matrix with $i$,$j$ element equal to the probability that a $\tau$-point mutation of the string $j_0 \dots j_{n-1}$ will yield the string $i_0 \dots i_{n-1}$.  Because all non-zero entries in this matrix are equal, it defines a Markov process with steady state probability vector $2^{-n}\vec{1}$ where $\vec{1}$ is the vector consisting of all ones.

\begin{thm}
\label{thm2.1}
The matrix $T_{n+1}(I,\tau)$ is iteratively generated from the $2 \times 2$ identity $I$ and anti-identity $I^*$ matrix by the recursion
\ben
\label{eq2.3}
T_{n+1}(I,\tau) =
\begin{pmatrix}
T_n(I,\tau) & T_n(I,\tau-1) \\
T_n(I,\tau-1) & T_n(I,\tau)
\end{pmatrix} \ .
\een
\end{thm}

\begin{pf}
To construct $T_{n+1}(I,\tau)$ write out a $2^{n+1}\times 2^{n+1}$ matrix with indices ranging from $0$ to $2^{n+1}-1$ in ascending numerical order.  In binary form the first $2^n$ of these indices will consist of strings of $n+1$ digits that begin with a $0$ while the second $2^n$ indices will consist of strings of $n+1$ digits that begin with a $1$.  This provides a partition of the matrix into four $2^n\times 2^n$ blocks.  Now consider a toggle of $\tau$ digits in the strings labelling the columns of the matrix.  If the first digit is not toggled then the effect on the remaining $n$ digits is identical to a $\tau$-point mutation on $n$-digit strings.  Thus, the first and fourth quadrant of the matrix contain $T_n(I,\tau)$.  On the other hand, if the first digit is toggled, the effect on the remaining $n$ digits is identical to a $(\tau-1)$-point mutation on the remaining $n$-digit strings.  Thus the second and third quadrants of the matrix contain $T_n(I,\tau-1)$.\qed
\end{pf}
 
\begin{cor}
\ben
\label{eq2.4}
\overline{T}_{n+1}(I,\tau) =
\begin{pmatrix}
\frac{n-\tau+1}{n}\overline{T}_n(I,\tau)   &   \frac{\tau}{n+1}\overline{T}_n(I,\tau-1) \\
\frac{\tau}{n+1}\overline{T}_n(I,\tau-1)   &   \frac{n-\tau+1}{n}\overline{T}_n(I,\tau)
\end{pmatrix}
\een
\end{cor}

\begin{thm}
\label{thm2.2}
For all $n$ and $\tau$ there exists a set of permutation matrices
\be
\left\{P_s(n,\tau) | 1\le s\le \tbinom{n}{\tau}\right\} \ ,
\ee
such that
\ben
\label{eq2.5}
T_n(I,\tau) = \sum_{s=1}^{\binom{n}{\tau}} P_s(n,\tau) \ .
\een
\end{thm}

\begin{pf}
By Theorem \ref{thm2.1},
\be
T_n(I,\tau) =
\begin{pmatrix}
T_{n-1}(I,\tau) & T_{n-1}(I,\tau-1) \\
T_{n-1}(I,\tau-1) & T_{n-1}(I,\tau)
\end{pmatrix} .
\ee
Suppose that there are sets of permutation matrices $\{Q_i\}$ and $\{{Q'}_i\}$ such that
\ben
\label{eq2.6}
\begin{array}{ccc}
T_{n-1}(I,\tau) &=& \sum\limits_{i=1}^{a} Q_i \\
T_{n-1}(I,\tau-1) &=& \sum\limits_{i=1}^{b} {Q'}_i
\end{array} \ .
\een
Since each row and column of $T_n(I,\tau)$ contains $\tbinom{n}{\tau}$ ones, the summation indices are $a=\tbinom{n-1}{\tau}$ and $b=\tbinom{n-1}{\tau-1}$. Thus, $T_n(I,\tau)$ can be written in the form
\ben
\label{eq2.7}
T_n(I,\tau) = 
\begin{pmatrix}
\sum\limits_{i=1}^{a} Q_i & 0 \\
0 & \sum\limits_{i=1}^{a} Q_i
\end{pmatrix} + 
\begin{pmatrix}
0 & \sum\limits_{i=1}^{b} {Q'}_i \\
\sum\limits_{i=1}^{b} {Q'_i} & 0
\end{pmatrix} \ .
\een

The first matrix defines a set of  permutations on $\{0,\dots,2^n-1\}$ in terms of the permutations $\{Q_i\}$ defined on $\{0,\dots,2^{n-1}-1\}$ by
\ben
\label{eq2.8}
x \in \{0,\dots,2^n-1\} \rightarrow
\begin{cases}
Q_i(x) & x \in \{0,\dots,2^{n-1}-1\} \\
Q_i(x-2^{n-1})+2^{n-1} & x \in \{2^{n-1},\dots,2^n-1\}
\end{cases} \ .
\een

The second matrix defines a set of $\tbinom{n-1}{\tau-1}$ permutations on $\{0,\dots,2^n-1\}$ in terms of the $\{{Q'}_i\}$ by
\ben
\label{eq2.9}
x \in \{0,\dots,2^n-1\} \rightarrow
\begin{cases}
{Q'}_i(x)+2^{n-1} & x \in \{0,\dots,2^{n-1}-1\} \\
{Q'}_i(x-2^{n-1}) & x \in \{2^{n-1},\dots,2^n-1\}
\end{cases} \ .
\een

Since $\tbinom{n-1}{\tau}$ + $\tbinom{n-1}{\tau-1}$ = $\tbinom{n}{\tau}$ this expresses $T_n(I,\tau)$ as a sum of $\tbinom{n}{\tau}$ permutations, so long as \eqref{eq2.6} is satisfied.  But for all $\tau > 0$, $T_{\tau}(I,\tau) = I^*$ while for $\tau = 0$ and all $n$, $T_n(I,0) = I$.  Since both $I$ and $I^*$ are permutation matrices the result follows by induction. \qed
 
Let $\mu$ be an $n$-digit binary string.  The parity $\pi(\mu)$ of $\mu$ is defined as $0$ if $\mu$ contains an even number of ones and $1$ if $\mu$ contains an odd number of ones.  A string will also be referred to as having even or odd parity in these two cases.  On this basis a partition of the state space $E_n$ is defined by $E_n = E_n^{(e)}\cup E_n^{(o)}$ where
\ben
\label{eq2.10}
\mu \in
\begin{cases}
E_n^{(e)} & \text{if $\pi(\mu) = 0$} \\
E_n^{(o)} & \text{if $\pi(\mu) = 1$}
\end{cases} \ .
\een

If $\tau$ is odd then a $\tau$-point mutation takes elements of $E_n^{(e)}$ to $E_n^{(o)}$ and vice versa while if $\tau$ is even it takes elements of $E_n^{(e)}$ to $E_n^{(e)}$ and elements of $E_n^{(o)}$ to $E_n^{(o)}$.  In addition, for $\tau > 0$ no element of $E_n$ will mutate to itself so the graph $H_n(\tau)$ contains no loops.  Since each mutation is reversible, however, the shortest cycle period in $H_n(\tau)$ is always two.
\end{pf}

\begin{lem}
\label{lem2.1}
Let $M_n$ be any $2^n\times2^n$ matrix with indices labelled in ascending order from 0 to $2^n-1$.  Then there is a permutation matrix $P_n$ such that the indices $i$,$j$ of $[P_n^{-1}M_nP_n]_{ij}$, when expressed in binary form, satisfy
\ben
\label{eq2.11}
\pi(i) = \pi(j) = 
\begin{cases}
0 & \text{$0 \le i$, $j \le 2^{n-1}-1$} \\
1 & \text{$2^{n-1} \le i$, $j \le 2^{n}-1$}
\end{cases} \ .
\een
In addition, if $\pi(i) \ne \pi(i')$ and $i < i'$ numerically then $i$ precedes $i'$ as an index.  (That is, both even and odd parity indices are given in ascending numerical order.)  Further, if $P_{n+1}$ is the permutation matrix that produces this ordering for $2^{n+1}\times2^{n+1}$ matrices then $P_{n+1}$ is obtained from $P_n$ as follows: write $P_n^{-1}$ in terms of two $2^{n-1}\times2^n$ matrices $Q$ and $R$ in the form
\ben
\label{eq2.12a}
P_n^{-1} = \begin{pmatrix} Q \\ R \end{pmatrix} .
\een
Then,
\ben
\label{eq2.12b}
P_{n+1}^{-1} = \begin{pmatrix} Q & 0 \\ 0 & R \\ R & 0 \\ 0 & Q \end{pmatrix} .
\een
\end{lem}

The index ordering resulting from application of Lemma \ref{lem2.1} will be called parity ordering.

\begin{lem}
\label{lem2.2}
With parity ordering of indices the matrix $T_n(I,\tau)$ takes the form
\ben
\label{eq2.13}
T_n(I,t) =
\begin{cases}
\begin{pmatrix} 0 & A \\ A & 0 \end{pmatrix} & \text{$t$ odd} \\
\begin{pmatrix} B & 0 \\ 0 & B \end{pmatrix} & \text{$t$ even}
\end{cases} \ .
\een
\end{lem}

\begin{pf}
Since $T_n(I,\tau)$ is symmetric, and odd toggles change the parity of a state while even toggles preserve it, $T_n(I,\tau)$ must at least have the form
\ben
\label{eq2.14}
T_n(I,t) =
\begin{cases}
\begin{pmatrix} 0 & A \\ A^T & 0 \end{pmatrix} & t \mathrm{\ odd} \\
\begin{pmatrix} B & 0 \\ 0 & C \end{pmatrix} & t \mathrm{\ even}
\end{cases} \ ,
\een
where $A$, $A^T$, $B$, and $C$ are all square matrices of the same size.  Let $j$ be the binary form of the $j$-th even index.  A $\tau$-point mutation of $j$ has the form $j + \eta$ where the $n$ digit binary string $\eta$ contains $\tau$ ones and addition is site-wise $\mmod(2)$.  For odd $\tau$, $j + \eta \in E_n^{(o)}$ while $j + \eta + \alpha_n \in E_n^{(e)}$, where $\alpha_n$ is the string $0\dots01$ with a single 1 in the $n$-th position.  But $j$ and $j + \alpha_n$ label corresponding columns of $A^T$ and $A$ while $j + \eta$ and $j + \eta + \alpha_n$ label corresponding rows of $A^T$ and $A$.  Thus $A^T = A$.  Likewise, if $\tau$ is even then $j$ and $j + \alpha_n$ label corresponding columns of $B$ and $C$ while $j + \eta$ and $j + \eta + \alpha_n$ label corresponding rows of $B$ and $C$, showing that $C = B$. \qed
\end{pf}

\begin{thm}
\label{thm2.3}
If $\tau$ is odd $T_n(I,\tau)$ is irreducible and bipartite.
\end{thm}

\begin{pf}
If $\tau$ is odd there can be no cycles in $H_n(I,\tau)$ having odd period since successive mutations will oscillate between $E_n^{(e)}$ and $E_n^{(o)}$.  Since period two cycles are possible, the index of imprimitivity of $T_n(I,\tau)$ is 2 and thus $T_n(I,\tau)$ is bipartite and irreducible. \qed
\end{pf}

\begin{lem}[{\cite[p.74]{ref32}}]
\label{lem2.3}
Let $M$ be an irreducible non-negative matrix of order $n$ with index of imprimitivity $k$ and let $m$ be a positive integer.  Then $M^m$ is irreducible if and only if $k$ and $m$ are relatively prime.
\end{lem}

Since for odd $\tau$ the index of imprimitivity of $T_n(I,\tau)$ is 2, this implies that if $\tau$ is odd $T_n^m(I,\tau)$ will be irreducible if and only if $m$ is odd.  Note also that Theorem \ref{thm2.3} implies the graph $H_n(\tau)$ is strongly connected if and only if $\tau$ is odd, while for even $\tau$ this graph is composed of two disjoint strongly connected components with respective vertex sets $E_n^{(e)}$ and $E_n^{(o)}$.

\section{State Transition Representations}
\label{sec3}
The space $E_n$ of $n$-digit binary strings with periodic boundary conditions is the state space for all binary valued cellular automata acting on cylinders of size $n$.  In Section \ref{sec2} the structure of the graph $H_n(\tau)$ was studied on this space.  This structure describes the results of point mutations.  The action of a CA rule on $E_n$ can also be described in terms of a graph.  This is usually referred to as the state transition diagram, but there are other diagrams that also provide information.

\subsection{State Transition Diagrams}
\begin{figure}
\begin{center}
\resizebox{\linewidth}{!}{\includegraphics{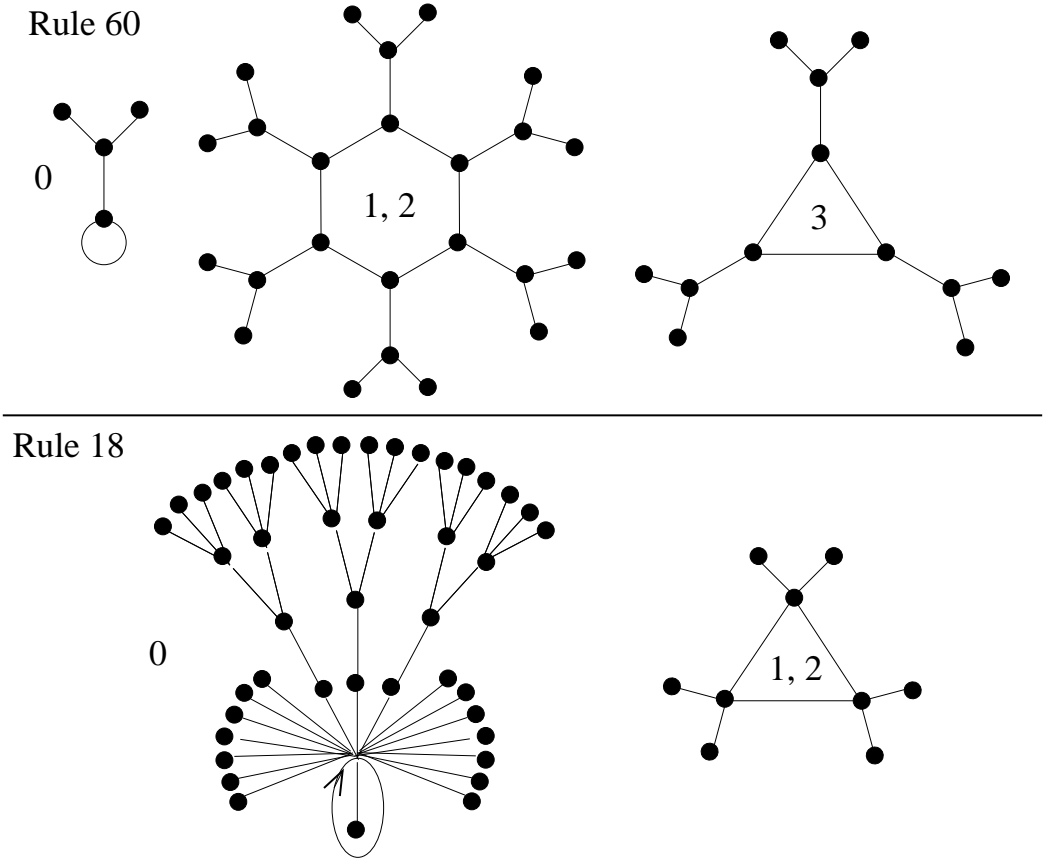}}
\caption{State transition diagrams for left justified rule $60$ and $18$ on a cylinder of size $6$. The numbers accompanying each diagram enumerate the cycle structures. The presence of more than one number indicates that several cycles have the same structure.}
\label{fig1}
\end{center}
\end{figure}

A cellular automaton rule $X$ acting on $E_n$ defines a directed graph $G(X)$ with vertex set $E_n$ and with an edge from $\mu$ to $\mu'$ if and only if $X(\mu) = \mu'$.  This is the standard state transition diagram (STD) for $X$ on $E_n$.  The structure and generation of this diagram has been extensively studied \cite{ref23,ref22}.  The STD partitions into disjoint subgraphs, each of which constitutes a basin of attraction.  Each such basin contains an attractor, either a cycle or fixed point, of the CA rule. Figure \ref{fig1} shows the STD's for the binary difference rule (rule 60) \cite{ref20} and the extensively studied rule 18 \cite{ref33} on a cylinder of size $6$.
 
A left-justified $k$-site CA rule $X$ is additive if it can be represented in the form
\ben
\label{eq3.1}
X = \sum\limits_{i=0}^{k-1} a_i \sigma^i \ ,
\een
where $\sigma$ is the left shift and the coefficients, $a_i$, take values in $\{0,1\}$.  Additive rules are particularly well behaved.  For example, \eqref{eq3.1} is equivalent to the condition $X(\mu+\mu') = X(\mu) + X(\mu')$ for all states $\mu$ and $\mu'$.  The state $\vec{0}$ consisting of all zeros is a fixed point for all additive rules and the state $\vec{1}$ consisting of all ones is a fixed point for those rules in which an odd number of the coefficients in \eqref{eq3.1} are not zero.

\begin{lem}[\cite{ref28}]
\label{lem3.1}
Let $X$ be the global transition rule for an additive cellular automaton.  Then the state transition diagram of $X$ acting on $E_n$ consists of cycles and fixed points with trees rooted at each state on a cycle and at each fixed point.  Further, all trees are topologically isomorphic to the tree rooted at the fixed point $\vec{0}$.
\end{lem}

If $X$ is an additive rule then its parity can be defined in terms of the representation (\ref{eq3.1}) by
\ben
\label{eq3.2}
\pi(X) =
\begin{cases}
0 & \text{an even number of the $a_i$ are $1$} \\
1 & \text{an odd number of the $a_i$ are $1$}
\end{cases} \ .
\een

Site-wise binary addition of an even number of binary strings always yields another string with even parity, while addition of an odd number of binary strings of even parity yields a string with even parity and addition of an odd number of binary strings of odd parity yields a string with odd parity.  Thus, we have:
\begin{lem}
\label{lem3.2}
Let $X$:$E_n \rightarrow E_n$ be an additive rule.
\begin{enumerate}
\item If $\pi(X) = 0$ then $X$ maps both $E_n^{(e)}$ and $E_n^{(o)}$ to $E_n^{(e)}$.
\item If $\pi(X) = 1$ then $X$ maps $E_n^{(e)}$ to $E_n^{(e)}$ and $E_n^{(o)}$ to $E_n^{(o)}$.
\end{enumerate}
\end{lem}

An immediate consequence of this lemma, together with \eqref{eq3.1}, is that if $X$ is an additive rule with even parity then no state with odd parity can have a predecessor in $E_n$.  Thus the states in $E_n^{(o)}$ must reside at the top of the trees in the STD.  Also, if $X$ has odd parity the STD must partition into two distinct components with $E_n^{(e)}$ and $E_n^{(o)}$ as their respective vertex sets.  In general, states without predecessors will be called peripheral and all other states will be called internal. 

Proof of the next theorem follows directly from Lemma \ref{lem3.2}.

\begin{thm}
\label{thm3.1}
Let $X$ be an additive rule and let $\{\mu(s) | 1 \le s \le p\}$ be a cycle of $X$ having period $p$.  Then all states on this cycle have the same parity.
\end{thm}

\begin{lem}
\label{lem3.3}
Let $X$:$E_n \rightarrow E_n$ be additive with $\pi(X) = 1$ and let $n$ be odd.  Then the components of the STD of $X$ corresponding to the vertex sets $E_n^{(e)}$ and $E_n^{(o)}$ are isomorphic.
\end{lem}

\begin{pf}
Let $\mu\in E_n^{(e)}$ and set $\mu' = \vec{1}+\mu$.  Since $n$ is odd, $\mu'\in E_n^{(o)}$ and $X(\mu') = X(\vec{1}) + X(\mu)$.  Further, $X(\vec{1}) = \vec{1}$ since it is the sum of an odd number of shifts of $\vec{1}$.  Thus for all $\mu\in E_n^{(e)}$ there is an element $\vec{1}+\mu \in E_n^{(o)}$ that maps in an identical manner under $X$ and vice versa. \qed
\end{pf}

Note that rule 150 acting on $E_4$ provides a counter-example to Lemma \ref{lem3.3} in the case that $n$ is even.  The catch is that for even $n$ both $\vec{0}$ and $\vec{1}$ are in $E_n^{(e)}$.

\begin{lem}
\label{lem3.4}
Let $X$ be an additive rule with $\pi(X) = 0$.  If the cylinder size $n$ is odd then the maximum tree height $h^*$ is equal to 1.
\end{lem}

\begin{pf}
With $X$ given by \eqref{eq3.1} the rule can be represented by the $n\times n$ circulant matrix $X_C = \circul_n(a_0,\dots,a_{k-1})$.  If $n$ is odd and $\pi(X) = 0$ this is a matrix of odd order with each row and column containing the same even number, say $2r$, of ones.  Thus, summation $\mmod(2)$ of the first $n-1$ rows of $X_C$ and addition, $\mmod(2)$, of this sum to the final row must yield a row of zeros.  But it is not possible, using elementary operations and $\mmod(2)$ arithmetic, to produce another row of zeros in this matrix.  To see this, consider the $i$-th column and the $(n-1)$-st row.  The entry in this position will be a 0 or a 1 and the $i$-th entry in the original $n$-th row will also have been a 0 or a 1.  There are therefore four cases.  The number of ones in the $i$-th column that lie above the final two rows are indicated in Table \ref{tab1} for each case.
\begin{table}
\begin{center}
\begin{tabular}{|c|c|c|c|}
\cline{3-4}
\multicolumn{2}{c|}{ } & \multicolumn{2}{|c|}{$(n-1)$-st row} \\
\cline{3-4}
\multicolumn{2}{c|}{ } & 0 & 1 \\
\hline
\raisebox{-1.5ex}[0pt]{$n$-th row} & 0 & 2r & 2r-1 \\
\cline{2-4}
           & 1 & 2r-1 & 2r-2 \\
\hline
\end{tabular}
\end{center}
\caption{The number of ones in the $i$-th column that lie above the final two rows.}
\label{tab1}
\end{table}

If a second row of zeros is to be possible, a linear combination of the first $n-2$ rows of $X_C$ must equal the value in the $(n-1)$-st row for each column.  This combination must have an odd number of ones in those columns in which the $(n-1)$-st row has a 1, and an even number of ones in the columns in which the $(n-1)$-st row has a 0.  Examination of Table \ref{tab1}, however, yields the parity of the number of rows required for each block of Table \ref{tab1}: both entries in the first column must be even and both in the second column must be odd.  This indicates that the only case in which the contradiction of requiring both an odd and an even number of rows in the sum will not occur is if the $(n-1)$-st row consists of all zeros or all ones.  The first case corresponds to the 0-rule and the second cannot occur since $n$ is odd while $\pi(X) = 0$. Thus, the nullity of the matrix $X_C$ is $\nu = 1$.  By a theorem of Martin, et al.\ \cite{ref28}, the in degree of fixed points or states on a cycle of an additive rule is $2^{\nu}$.  Thus, for the cases in question the in degree of $\vec{0}$ is two and since $X(\vec{0}) = \vec{0}$ and $X(\vec{1}) = \vec{0}$ this means that $\vec{1}$ is the only non-trivial predecessor of $\vec{0}$.  Since all trees are topologically isomorphic to the tree rooted at $\vec{0}$ the proof is done. \qed
\end{pf}

For additive rules with $\pi(X) = 1$ the situation is more complicated.  If $\pi(X) = 0$ the minimum possible height for trees is 1 since elements of $E_n^{(o)}$ have no predecessors.  If $\pi(X) = 1$ this is not the case and the minimum possible height is 0.  This occurs trivially for $X = \sigma^k$ for any $k$, but other cases for which $h^* = 0$ also exist as indicated by the next theorem.

\begin{thm}[\cite{ref20}]
\label{thm3.2}
The maximum tree height for the rule $I + \sigma + \sigma^2$ (rule 150, left justified) is 0 unless $n = 0 \pmod 3$.
\end{thm}

\begin{conj}
Let $X$ be an additive rule and let $\kappa(X)$ denote the number of non-zero coefficients in the expression (\ref{eq3.1}).  If $\pi(X) = 1$ then $h^* = 0$ unless $\kappa(X)|n$.
\end{conj}

In describing transitions between attractor basins there are several ways to represent the set of basins for any given rule.  In this paper it is assumed that mutations occur at a rate much faster than the time scale of rule iteration.  If it is assumed that mutations occur on a time scale that is much slower than rule iterations (e.g., as is the case in the computation of ``metagraphs'' in discrete dynamics lab \cite{ref34}) then only the attractors are relevant.

\subsection{The State-Basin Representation}
If $B_{\alpha}$ is an attractor basin for a CA rule $X$ and $\mu\in B_{\alpha}$ then the height $h$ of $\mu$ above the attractor is 0 if $\mu$ is on the attractor and is the minimum integer $h$ such that $X^h(\mu)$ lies on the attractor otherwise.

If $\{B_{\alpha}| 0\le\alpha\le a\}$ is the set of attractor basins for a CA rule $X$ then $E_n$ can be partitioned into disjoint classes labelled $\alpha$:$h(\alpha)$ where $\alpha$ indicates a specific attractor and $h(\alpha)$ is a height above the attractor $\alpha$ in the basin $B_{\alpha}$.  Thus, each class consists of those states that are at a specified height above a given attractor.  Conventionally, if $\vec{0}$ is a fixed point the corresponding basin will be denoted $B_0$.  This defines what will be called the state-basin partition of $E_n$.

Classes in the state-basin partition are taken as the vertices of a graph $H_n(X,\tau)$ with an edge connecting two vertices for each $\tau$-point mutation that takes a state in the state-basin class labelling one vertex to the state-basin class labelling the other vertex.  The matrix $T_n(X,\tau)$ is defined as the weighted adjacency matrix of this graph.  If $X$ is additive, $T_n(X,\tau)$ will have size $N\times N$ with $N=\sum_{\alpha}[h^*(\alpha)+1]$ where $h^*(\alpha)$ is the maximum tree height for basin $B_{\alpha}$.

Clearly $T_n(X,\tau)$ is symmetric and $T_n(I,\tau)$ is the special case in which $X$ is the identity rule.  $T_n(X,\tau)$ can be directly obtained from $T_n(I,\tau)$ by summing over the rows and columns corresponding to each state-basin class.  Unfortunately there are no general equivalents to Theorems \ref{thm2.1}, \ref{thm2.2}, and \ref{thm2.3} that are valid for arbitrary rules.

As with $T_n(I,\tau)$, a probability matrix $\overline{T}_n(X,\tau)$ is defined by dividing each column of $T_n(X,\tau)$ by its sum.  The state-basin classes for the binary difference rule $D$, and $\overline{T}_n(D,1)$ are given in Appendix \ref{ap2} for a cylinder of size 6.

The value of $[T_n(X,\tau)]_{ij}$ gives the number of distinct paths in $H_n(X,\tau)$ from vertex $j$ to vertex $i$ and the characteristic polynomial of $T_n(X,\tau)$ provides information on the number of cycles in this graph.

\begin{lem}[{\cite[p.76]{ref32}}]
\label{lem3.5}
Let $\phi(\lambda) = \lambda_n + c_1\lambda^{n-1} + c_2\lambda^{n-2} + \dots + c_n$ be the characteristic polynomial of a matrix $M$ which is the adjacency matrix of a graph $G$.  Then
\begin{enumerate}
\item The coefficient $c_r$ of $\lambda^{n-r}$ is the sum of the determinants of all principle minors of $M$ of size $r$.
\item The value of $|c_r|$ equals the number of cycles of $G$ with periods that sum to $r$.
\end{enumerate}
\end{lem}

\begin{lem}[{\cite[p.77]{ref32}}]
\label{lem3.6}
Let $M$ be an $r$-cyclic $n\times n$ matrix with $M^r$ having block diagonal form
\be
M^r =
\begin{pmatrix}
B_1 & 0 & 0 & \cdots & 0 \\
0 & B_2 & 0 & \cdots & 0 \\
0 & 0 & B_3 & \cdots & 0 \\
\vdots & \vdots & \vdots & \ddots & \vdots\\
0 & 0 & 0 & \cdots & B_r
\end{pmatrix} ,
\ee
then there exists a monic polynomial $f(\lambda)$ and non-negative integers $p_1,\dots,p_r$ such that
\begin{enumerate}
\item $f(0) \ne 0$.
\item For all $i \le r$ the characteristic polynomial of $B_i$ is $\lambda^{p_i}f(\lambda)$.
\item The characteristic polynomial of $M$ is $\lambda^{p_1+\dots+p_r}f(\lambda^r)$.
\end{enumerate}
\end{lem}

Since the index of imprimitivity of the matrix in this lemma is $r$, it is $r$-periodic and every cycle must have period divisible by $r$.  Thus, the only non-zero coefficients in the characteristic polynomial are $c_{kr}$ for $0 \le k \le \left\lceil\frac{n}{r}\right\rceil$.  On this basis, if $\tau$ is odd and $T_n(X,\tau)$ is bipartite, the characteristic polynomial of $T_n(X,\tau)$ will have the form
\ben
\label{eq3.3}
\phi(\lambda) = \lambda^k + \sum\limits_{i=1}^{\frac{k}{2}}c_i \lambda^{k-2i} \ .
\een

\subsection{Shift-Basin and ACS Representations}
Every cellular automata rule commutes with the shift operator.  This leads to a close relation between the state transition diagram and an equivalent diagram defined in terms of shift cycles.  This will be called the shift-basin diagram (SBD).  The state space $E_n$ is first partitioned into shift equivalence classes with states $\mu$ and $\mu'$ belonging to the same shift class if and only if for some $r$ it is true that $\mu' = \sigma^r(\mu)$ where $[\sigma(\mu)]_i = \mu_{i+1}$.  The set of shift classes, $S(n) = \{S_r(n)\}$ is taken as a new state space and the rule $X$:$E_n\rightarrow E_n$ induces a mapping $X^*$:$S(n)\rightarrow S(n)$ by taking $X^*(S_j(n)) = S_i(n)$ if, for some $\mu\in S_j(n)$, $X(\mu)\in S_i(n)$.  The shift-basin diagram (SBD) is then defined as the state transition diagram of the map $X^*$.

\begin{figure}
\begin{center}
\resizebox{\linewidth}{!}{\includegraphics{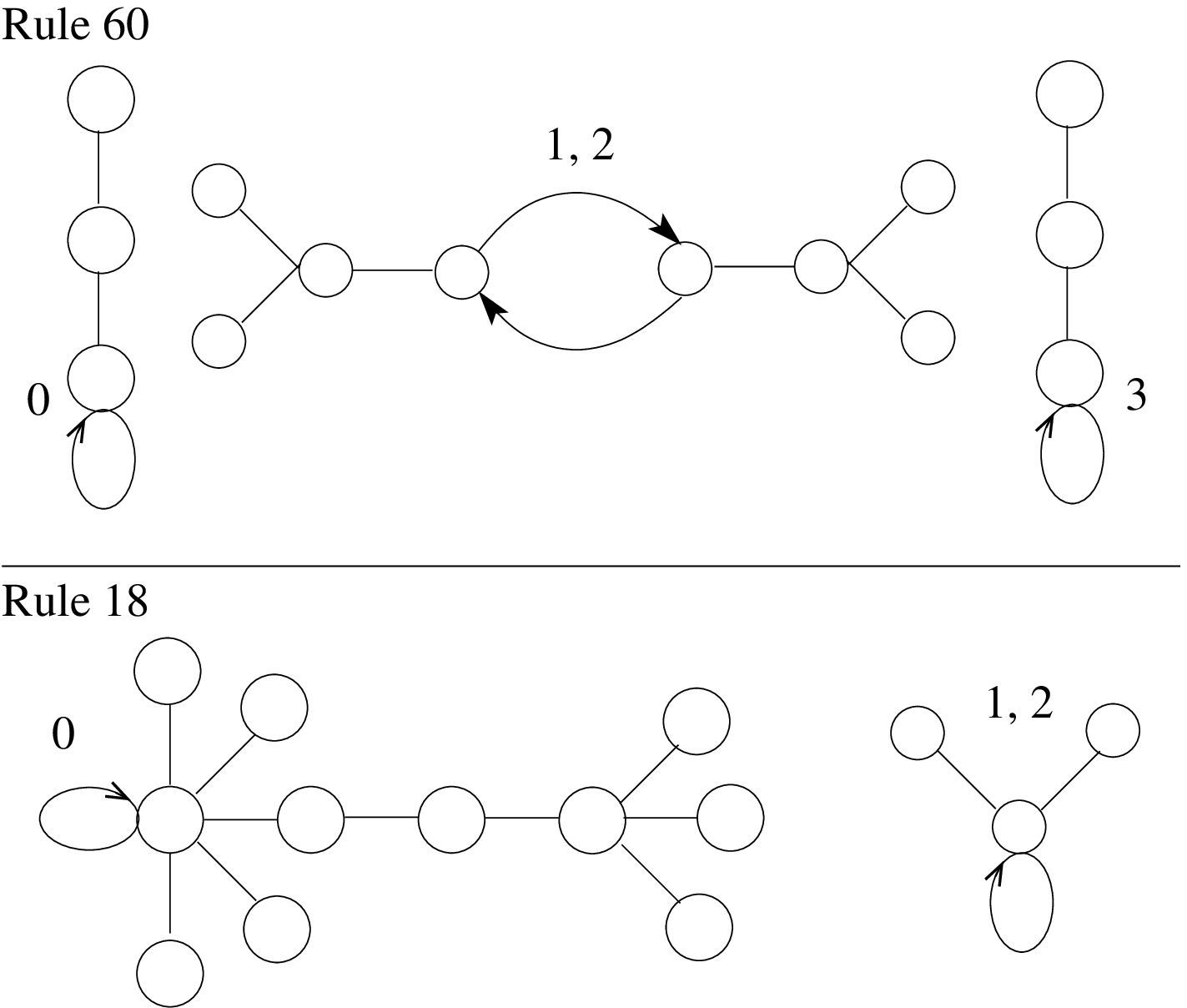}}
\caption{Shift-Basin Diagrams for left justified rules $60$ and $18$ on a cylinder of size $6$. The numbers accompanying each diagram enumerate the cycle structures. The presence of more than one number indicates that several cycles have the same structure.}
\label{fig2}
\end{center}
\end{figure}

If a cycle of $X$:$E_n\rightarrow E_n$ is a shift cycle it appears as a fixed point of $X^*$.  On the other hand, if a cycle of $X$:$E_n\rightarrow E_n$ consists of states drawn from $m$ distinct shift cycles then this appears as a period $m$ cycle of $X^*$.  Figure \ref{fig2} shows the SBD's for $n = 6$ for the binary difference rule and for rule 18.  These can be compared to the STD's of Figure \ref{fig1}.  The result on topological isomorphism between trees for additive CA's does not carry over to the shift-basin representation, as can be seen from Figure \ref{fig2}.

The following theorem relating the state-basin and shift-basin diagrams is a version of a result of Jen \cite{ref31}:
\begin{thm}
\label{thm3.3}
For a given $n$ let $S = \{S_i(n)|1 \le i \le m\}$ be an $m$-cycle of $X^*$:$S(n)\rightarrow S(n)$ and let $r$ be the smallest integer for which $X^m(\mu) = \sigma^{-r}(\mu)$ when $\mu$ lies in a shift class on this cycle.  Then $\mu$ lies on a cycle of $X$:$E_n\rightarrow E_n$ having period $ms$ where $s$ is the smallest integer such that $rs \equiv 0 \pmod n$.
\end{thm}

\begin{pf}
Let $\{c_i| 1 \le i \le m\}$ be an $m$-cycle of $X^*$.  Then each $c_i$ is a shift class in $E_n$ and since ${X^{*}}^m(c_i) = c_i$ for each state $\mu\in c_i$ there is a smallest integer $r < n$ such that $X^m(\mu) = \sigma^{-r}(\mu)$.  Thus, if $s$ is the smallest integer for which $rs = 0 \pmod n$ then $X^{ms}(\mu) = \mu$ and this is not true for $X^k(\mu)$ for any $k < ms$. \qed
\end{pf}

Another form of representation is based on the idea of an autocatalytic set (ACS) in a graph.  The vertices of a graph are first partitioned into two classes, those with in degree 0, and all others.  The set of vertices with in degree 0 form the periphery of the graph.  The set of all graph vertices is then partitioned into connected subgraphs and each such subgraph, excluding its peripheral vertices, is an ACS.  ``An autocatalytic set (ACS) is a subgraph, each of whose nodes has at least one incoming link from a node belonging to the same subgraph.'' \cite{ref15}.  An ACS basin representation can be constructed from either STD or SBD graphs.  In either case, the ACS basin representation may differentially mix parities from state representation or shift representation categories.

Each ACS and its corresponding peripheral set now becomes a pair of equivalence classes, and these classes again can be taken as forming a state space which will be denoted either $C_n$ or $C_n^*$ depending on whether it is defined from the STD or SBD diagram.  In either case the rule $X$ defines a map $X'$:$C_n\rightarrow C_n$ or $X''$:$C_n^*\rightarrow C_n^*$, and in both of these cases matrices $T_n(X',\tau)$, $\overline{T}_n(X',\tau)$, $T_n(X'',\tau)$, and $\overline{T}(X'',\tau)$ can be defined.

The shift-basin and ACS representations are useful since the associated matrices are smaller than that for the state-basin representation.  On the other hand, the state-basin representation will be more useful if probability distributions on attractor basins are used to model discrete potential wells.  In this paper the main focus will be on the state and shift representations.

\section{Properties of $T_n(X,\tau)$ and $\overline{T}_n(X,\tau)$}
A number of characteristics of the matrices $T_n(X,\tau)$ and $\overline{T}_n(X,\tau)$ can be determined, especially if the rule $X$ is additive.  For additive rules, all trees are topologically identical to the tree rooted at $\vec{0}$, providing a particularly nice link between the form of $T_n(X,\tau)$ and the structure of the state transition diagram.

Let $X$ be an additive rule with the state $\vec{0}$ having in degree $d$ and maximum tree height $h^*$.  Further, suppose that all trees are balanced, i.e., that all branchings are topologically identical and following any branch from $\vec{0}$ upward eventually reaches a height of $h^*$.  Then each attractor state has $d-1$ predecessors of height $1$ while all internal states not on an attractor have $d$ predecessors.  Thus, each rooted tree will contain $d^{h^*}$ states.  Of these, $d^{h^*-1}$ will be interior states and $d^{h^*-1}(d-1)$ will be peripheral states.

To go further, additional assumptions about the number of peripheral states will be required.  In addition, it is useful to choose the indexing of $T_n(X,\tau)$ so as to put this matrix into a simple form.  If $\pi(X) = 0$  interior/peripheral indexing is chosen, with indices representing classes of even parity that are composed of  peripheral states placed to the right of indices representing classes of even parity that are composed of interior states.  If $\pi(X) = 1$ parity indexing will be used.  Note that if $\pi(X) = 0$ and all even states have predecessors then interior/peripheral and parity indexing are the same.  Three simple cases will be considered.

\begin{case}: $\pi(X) = 0$ and the set of peripheral states is just the set of odd parity states. Since each tree contains $d^{h^*-1}(d-1)$ peripheral states this means that the total number of trees for a cylinder of size $n$ is given by
\ben
\label{eq4.1}
\frac{2^{n-1}}{d^{h^*-1} (d-1)} \ .
\een
Each tree is rooted at a fixed point or on a cycle, so (\ref{eq4.1}) is also the number of points that are fixed points or lie on cycles of $X$.  Since this number must be an integer, there will be constraints on the possible values of $n$ and $d$.  Suppose that $d = 2^r m$ where $m$ is odd and $r < n-1$.  Then \eqref{eq4.1} becomes
\be
\frac{2^{n-r(h^*-1)-1}}{(2^rm-1)m^{h^*-1}} \ ,
\ee
but $m$ is odd, hence both terms in the denominator of this expression are odd and it cannot be an integer unless $m$ and $r$ both equal $1$.  In that case $d = 2$ and the number of trees given by (\ref{eq4.1}) becomes $2^{n-h^*}$.

\begin{conj}
The only cellular automata rules satisfying the above conditions are multiples of the binary difference rule (rule 60) by powers of the shift.
\end{conj}

Under these conditions, the number of states at height $h$ in a tree is $2^{h-1}$ ($h \ge 1$) and the total number of states at height $h$ is $2^{n+h-h^*-1}$.  Let $N_p$ be the number of attractors of period $p$, including $p = 1$ for fixed points.  Then
\ben
\label{eq4.2}
\sum\limits_p = p N_p = 2^{n-h^*}
\een
and the total number of classes appearing in the definition of $T_n(X,\tau)$ is
\ben
\label{eq4.3}
\left( h^* + 1 \right)\ \sum\limits_p N_p
\een
of which $\sum_p N_p$ will be peripheral classes.  Since in this case interior/periphery and parity indexing are identical, $T_n(X,\tau)$ will have the form
\ben
\label{eq4.4}
T_n(X,\tau) =
\begin{cases}
\begin{pmatrix} 0 & A \\ A^T & 0 \end{pmatrix} & \text{$\tau$ odd} \\
\begin{pmatrix} B & 0 \\ 0 & C \end{pmatrix} & \text{$\tau$ even}
\end{cases} \ ,
\een
where the sizes of the matrices $A$, $B$, and $C$ are respectively $\left(h\cdot\sum_p N_p\right)\times\left(\sum_p N_p\right)$, $\left(h\cdot\sum_p N_p\right)\times\left(h\cdot\sum_p N_p\right)$, and $\left(\sum_p N_p\right)\times\left(\sum_p N_p\right)$.
\end{case}

\begin{case}: $\pi(X) = 0$ and the set of peripheral states contains $E_n^{(o)}$ and half of the states in $E_n^{(e)}$. Under these assumptions, with $d = 2^r m$, \eqref{eq4.1} becomes
\ben
\label{eq4.5}
\frac{3\cdot2^{n-r(h^*-1)-2}}{(2^r m-1) m^{h^*-1}} \ ,
\een
and this will be an integer if and only if $m = 1$ and $r = 2$.  In this case $d = 4$ and the total number of trees will be $2^{n-2h^*}$.  The well-known rule 90 is included under this case.

The number of states at height $h$ in each tree will be $4^{h-1}3 = 2^{2(h-1)}3$ and the equivalent of \eqref{eq4.2} is
\ben
\label{eq4.6}
\sum\limits_p = p N_p = 2^{n-2h^*} \ .
\een

In this case interior/periphery indexing differs from parity indexing.  If all peripheral classes contain only states of the same parity then use of parity indexing puts $T_n(X,\tau)$ into the form of \eqref{eq4.4}.  In general, however, peripheral classes will be of mixed parity and interior/peripheral indexing gives the form of $T_n(X,\tau)$ as
\ben
\label{eq4.7}
T_n(X,\tau) =
\begin{cases}
\begin{pmatrix} 0 & A \\ A^T & D \end{pmatrix} & \text{$\tau$ odd} \\
\begin{pmatrix} B & G \\ 0 & C \end{pmatrix} & \text{$\tau$ even}
\end{cases} \ ,
\een
where $A$, $D$, and $B$ have respective sizes $\left(h\cdot\sum_p N_p\right)\times\left(\sum_p N_p\right)$, $\left(\sum_p N_p\right)\times\left(\sum_p N_p\right)$, and $\left(h\cdot\sum_p N_p\right)\times\left(h\cdot\sum_p N_p\right)$, while $G$ is the same size as $A$ and $C$ the same size as $D$.  Note, however, that despite the form shown, the matrix $T_n(X,\tau)$ is always symmetric.
\end{case}

\begin{case}: $\pi(X) = 1$. By Lemma \ref{lem3.3}, if $n$ is odd then $T_n(X,\tau)$ in parity indexing must have the form
\ben
\label{eq4.8}
T_n(X,\tau) =
\begin{cases}
\begin{pmatrix} 0 & A \\ A & 0 \end{pmatrix} & \text{$\tau$ odd} \\
\begin{pmatrix} B & 0 \\ 0 & C \end{pmatrix} & \text{$\tau$ even}
\end{cases} \ ,
\een
where $A$, $B$, and $C$ are all square matrices of size $(h^*+1)/2\cdot\sum_p N_p$.  Lemma \ref{lem3.3} also implies that $N_p$ is even.

If $n$ is even then $T_n(X,\tau)$ has the form of (\ref{eq4.4}) but the size of the matrices $A$ and $B$ is now $\left((h^*+1)\sum_p N_p^{(e)}\right)\times\left((h^*+1)\sum_p N_p^{(o)}\right)$, while C is size $\left((h^*+1)\sum_p N_p^{(o)}\right)\times\left((h^*+1)\sum_p N_p^{(o)}\right)$.  Here $N_p^{(e)}$ and $N_p^{(o)}$ are the number of attractors of period $p$ with even or odd parity.
\end{case}

By Theorem \ref{thm3.1}, all states on an attractor of an additive rule will have the same parity.  Thus, each of the classes $\alpha$:$0$ for $0 \le \alpha \le N_p-1$ will consist of states having the same parity.  This is not generally true for non-additive rules, nor is it true in general for classes $\alpha$:$h$ for $h \ge 1$.  Thus, with appropriately chosen indexing, the matrix $T_n(X,\tau)$ for an additive rule can always be at least put into the form of \eqref{eq4.7}, although the sizes of the submatrices involved may vary.

What is more interesting, however, is the probability matrix $\overline{T}_n(X,\tau)$ for which the form of (\ref{eq4.7}) becomes
\ben
\label{eq4.9}
\overline{T}_n(X,\tau) =
\begin{cases}
\begin{pmatrix} 0 & A \\ B & C \end{pmatrix} & \text{$\tau$ odd} \\
\begin{pmatrix} D & G \\ 0 & M \end{pmatrix} & \text{$\tau$ even}
\end{cases} \ .
\een

\begin{lem}
\label{lem4.1}
Let $\overline{T}$ be any matrix with the form (\ref{eq4.9}) with $A$ having size $n\times m$ and $C$ having size $m\times m$.  Then the characteristic equation for $\overline{T}$ is obtained from
\ben
\label{eq4.10}
\begin{array}{cccc}
\lambda^{|n-m|}\left|\lambda^2I-\lambda C - BA \right| & = & 0, & \tau \mathrm{\ odd} \\
\left|\lambda I - D\right|\cdot\left|\lambda I-M\right| & = & 0, & \tau \mathrm{\ even}
\end{array} \ .
\een
\end{lem}

In some cases the term in \eqref{eq4.10} with $\tau$ odd can be factored.

For the matrices $T_n(I,\tau)$ and $\overline{T}_n(I,\tau)$ completely general results are available.

\begin{thm}
\label{thm4.1}
The eigenvalues of $T_n(I,\tau)$ and $\overline{T}_n(I,\tau)$ are, respectively, the eigenvalues of the matrices $T_{n-1}(I,\tau) \pm T_{n-1}(I,\tau-1)$ and $(n-\tau)/(n-1)\cdot\overline{T}_{n-1}(I,\tau) \pm \tau/n\cdot\overline{T}_{n-1}(I,\tau-1)$.
\end{thm}

\begin{pf}
By \eqref{eq2.3} the characteristic equation $|\lambda I - T_n(X,\tau)| = 0$ can be written as
\be
\left|\begin{array}{cc}
\lambda I - T_{n-1}(X,\tau) & -T_{n-1}(X,\tau-1) \\
-T_{n-1}(X,\tau-1) & \lambda I - T_{n-1}(X,\tau)
\end{array}\right| = 0 \ ,
\ee
which becomes
\be
\left| \lambda^2 I - 2\lambda T_{n-1}(I,\tau) + T^2_{n-1}(I,\tau) - T^2_{n-1}(I,\tau-1)\right| = 0 \ ,
\ee
or
\be
\left|\lambda I-\left[2T_{n-1}(I,\tau)+T_{n-1}(I,\tau-1)\right]\right|\cdot\left|\lambda I-\left[2T_{n-1}(I,\tau)+T_{n-1}(I,\tau-1)\right]\right| = 0 \ .
\ee

The result for $\overline{T}_n(I,\tau)$ follows from similar calculations based on \eqref{eq2.4}. \qed
\end{pf}

\begin{cor}
The eigenvalues $\lambda(n+1)$ and $\overline{\lambda}(n+1)$ of $T_{n+1}(I,\tau)$ and $\overline{T}_{n+1}(I,\tau)$ are given in terms of the eigenvalues $\lambda(n)$ and $\overline{\lambda}(n)$ of $T_n(I,\tau)$ and $\overline{T}_n(I,\tau)$ by
\ben
\label{eq4.11}
\begin{array}{ccc}
\lambda(n+1) &=& \lambda(n) \pm 1 \\
\overline{\lambda}(n+1) &=& \frac{n\overline{\lambda}(n) \pm 1}{n + 1}
\end{array} \ .
\een
\end{cor}

For the matrices $T_n(X,\tau)$ and $\overline{T}_n(X,\tau)$ the case is more complex, and only a few results are available.

\begin{thm}
\label{thm4.2}
Let $D$ represent the global operator for elementary rule 60 and let $n = 2^k$.  Then $\overline{T}_n(D,1)$ has the form
\be
\overline{T}_{2^k}(D,1) = \begin{pmatrix} 0 & A \\ 1 & 0 \end{pmatrix} ,
\ee
where $A$ is a $2^k\times1$ column with entries
\ben
\label{eq4.12}
A_h =
\begin{cases}
\frac{1}{2^{k+1}} & h = 0 \\
\frac{2^h}{2^{k+2}} & 1 \le h \le 2^k-1
\end{cases} \ ,
\een
and the $1$ in the lower left block represents a row consisting of $2^k$ ones.  The characteristic equation of $\overline{T}_n(D,1)$ is
\ben
\label{eq4.13}
\lambda^{2^k-1} \left( \lambda^2 - 1 \right) = 0 \ .
\een
\end{thm}

\begin{pf}
The STD for rule 60 with $n = 2^k$ consists of a single tree rooted at $\vec{0}$ with maximum height $h^* = 2^k$.  Further, by Lemma \ref{lem3.2}, all of the odd parity states reside at the top of this tree.  Thus, there is a single odd class, labelled $0$:$2^k$ and there are $2^k$ even parity classes labelled $0$:$h$ for $0 \le h \le 2^k-1$.  Further, for $h \ge 1$ each of the classes $0$:$h$ contains $2^h-1$ states while the class $0$:$0$ contains only a single member.  Each member of the interior classes is even and hence with $\tau = 1$ mutates to an odd state in the class $0$:$2^k$ a total of $n = 2^k$ times (one mutation for each digit in the state).  Thus the ($0$:$2^k$,$0$:$h$) entry of the matrix $T_{2^k}(D,1)$ is $2^{k+h-1}$ while the ($0$:$h'$,$0$:$h$) entries will be $0$ for $0 \le h$, $h' < 2^k$.  The ($0$:$2^k$,$0$:$2^k$) entry is $0$, and since the matrix must be symmetric, the ($0$:$h$,$0$:$2^k$) entries are also $2^{k+h-1}$.  Hence the column sum of the final column of the matrix is
\be
2^k + \sum\limits_{h=1}^{2^k-1} 2^{k+h-1} = 2^k\left[1+\sum\limits_{h=1}^{2^k-1}2^{h-1}\right] = 2^{2k+1} \ .
\ee

Dividing each element of the final column by this sum yields the form given in \eqref{eq4.12} while division of each element of the final row of the matrix by the corresponding column sum yields 1.  The characteristic equation then follows immediately as an application of Lemma \ref{lem4.1}. \qed
\end{pf}

Another result following from Lemma \ref{lem4.1} generalizes this theorem:
\begin{thm}
\label{thm4.3}
Let $A$ be an $m\times m$ probability matrix and let $\{p_i|1\le i\le r,\ 0\le p_i\le1\}$ be a set of non-negative numbers with $\sum_{i=1}^r p_i \le 1$.  Define an $(r+1)m\times (r+1)m$ probability matrix $\overline{T}$ by
\ben
\label{eq4.14}
\overline{T} =
\begin{pmatrix}
0 & \dots & \dots & 0 & p_1 A \\
0 & \dots & \dots & 0 & p_2 A \\
\vdots &  &    & \vdots & \vdots \\
0 & \dots & \dots & 0 & p_r A \\
A & \dots & \dots & A & \left( 1 - \sum\limits_{i=1}^{r} p_i\right) A
\end{pmatrix} \ .
\een
Then the characteristic equation of $\overline{T}$ is obtained from
\ben
\label{eq4.15}
\lambda^{(r-1)m} \left|\lambda I - A\right| \cdot \left|\lambda I+\sum\limits_{i=1}^{r}p_i A \right| = 0 \ .
\een
\end{thm}

An example of Theorem \ref{thm4.3} is given by rule $90$ on a cylinder of size 6.  Taking $\delta$ as the global operator for rule $90$, the matrix $\overline{T}(\delta,1)$ is
\ben
\label{eq4.16}
\overline{T}_6(\delta,1) = \begin{pmatrix}0 & \frac{1}{3}A \\ A & \frac{2}{3}A \end{pmatrix} \ ,
\een
with
\ben
\label{eq4.17}
A =
\begin{pmatrix}
0           & 0           & \frac{1}{6} & \frac{1}{6} & 0           & 0           \\
0           & 0           & \frac{1}{6} & \frac{1}{6} & \frac{1}{3} & \frac{1}{3} \\
\frac{1}{2} & \frac{1}{6} & \frac{1}{3} & 0           & \frac{1}{6} & \frac{1}{6} \\
\frac{1}{2} & \frac{1}{6} & 0           & \frac{1}{3} & \frac{1}{6} & \frac{1}{6} \\
0           & \frac{1}{3} & \frac{1}{6} & \frac{1}{6} & 0           & \frac{1}{3} \\
0           & \frac{1}{3} & \frac{1}{6} & \frac{1}{6} & \frac{1}{3} & 0
\end{pmatrix} \ .
\een

The characteristic polynomial of $\overline{T}_6(\delta,1)$ is obtained from $|\lambda I-A|\cdot|\lambda I+A/3| = 0$.

Appendix \ref{ap1} lists the characteristic polynomials of $\overline{T}_n(X,1)$ for a number of additive and non-additive rules for varying cylinder sizes.

The matrix $T_n(X,\tau)$ has an immediate interpretation as the weighted adjacency matrix of the graph $H_n(X,\tau)$ with vertices labelled by the classes $\alpha$:$h(\alpha)$ and the edges between any pair of vertices weighted by the number of $\tau$-point mutations that take elements from the class labelling one vertex to that labelling the other and vice versa.  Thus the ($\alpha$:$h(\alpha)$, $\beta$:$h(\beta)$) element of $T_n(X,\tau)$ is the probability that a $\tau$-point mutation of the class $\beta$:$h(\beta)$ will be in the class $\alpha$:$h(\alpha)$.  While $T_n(X,\tau)$ is necessarily symmetric, this is not in general true for $\overline{T}_n(X,\tau)$.

An intuitive understanding of the spectra of both $T_n(X,\tau)$ and $\overline{T}_n(X,\tau)$ is obtained by relating the eigenvalues of these matrices to the solvability conditions for a system of linear equations connected with the graph $H_n(X,\tau)$.  Label each vertex of this graph by a variable $x_i$.  Then the question of finding values $y_i$, not all zero, of these variables such that each $y_i$ is proportional, with the same constant of proportionality, to the sum of all values $y_j$ such that there is an edge connecting vertex $j$ to vertex $i$ is equivalent to solving the homogeneous system of equations
\ben
\label{eq4.18}
\lambda x_i = \sum\limits_{j\rightarrow i} \left[ T_n(X,\tau)\right]_{ij} x_j \mathrm{\ or\ equivalently\ } \lambda \vec{x} = T_n(X,\tau) \vec{x} \ ,
\een
hence the possible proportionality factors $\lambda$ are just the eigenvalues of $T_n(X,\tau)$.

If, instead, each value $y_i$ is required to be proportional to the mean value of all $y_j$ such that there is an edge connecting vertex $j$ to vertex $i$ the corresponding set of linear equations to be satisfied becomes
\ben
\label{eq4.19}
\lambda x_i = \frac{1}{d_i} \sum\limits_{j\rightarrow i} \left[T_n(X,\tau)\right]_{ij} x_j \mathrm{\ or\ } \lambda \vec{x} = T_n(X,\tau) D^{-1} \vec{x} \ ,
\een
where $d_i$ is the in degree of vertex $i$, namely
\ben
\label{eq4.20}
d_i = \sum\limits_j \left[ T_n(X,\tau)\right]_{ij} \mathrm{\ and\ } D = \mathrm{diag}(d_i) \ .
\een

With these definitions it is clear that $\overline{T}_n(X,\tau) = T_n(X,\tau)D^{-1}$ so that the eigenvalues of $\overline{T}_n(X,\tau)$ are the possible proportionality factors for which the numerical values of each $x_i$ are proportional to the mean of all $x_j$ values at vertices $j$ having an edge connecting them to vertex $i$.  If the characteristic polynomials of $T_n(X,\tau)$ and $\overline{T}_n(X,\tau)$ are written respectively as $P(\lambda) = \lambda^n + a_1\lambda^{n-1} + \dots + a_n$ and $Q(\lambda) = \lambda^n + q_1\lambda^{n-1} + \dots + q_n$. Then the coefficients of these polynomials are \cite{ref35}:
\ben
\label{eq4.21}
\begin{array}{lll}
a_i &=& \sum\limits_{\stackrel{L\in \Lambda_i}{i=1,\dots,n}} \left( -1 \right)^{p(L)} \\
q_i &=& \sum\limits_{U\in Y_i} \left( -1 \right)^{p(U)} \frac{2^{c(U)}}{\prod\limits_{h\in V(U)} d_h}
\end{array} \ ,
\een
where $\Lambda_i$ is the set of linear directed subgraphs on $i$ vertices, $p(L)$ is the number of cycles in the linear directed subgraph $L$, $Y_i$ is the set of basic figures of size $i$, $p(U)$ is the number of components in $U$, $c(U)$ is the number of circuits in $U$, and $V(U)$ is the vertex set of $U$.

Since $\overline{T}_n(X,\tau)$ is a probability matrix it has maximum eigenvalue $1$, corresponding to the case in which the numerical values of each variable $x_i$ in \eqref{eq4.19} is equal to the mean of all values $x_j$ such that there is an edge connecting vertex $j$ to vertex $i$.

\begin{thm}
\label{thm4.4}
Let the matrix $\overline{T}_n(X,\tau)$ be defined in state, shift, or ACS representation.  Let $\{p(\tau)\in[0,1]|0\le\tau\le k\}$ be a set of non-negative numbers such that $\sum_{\tau=1}^k p(\tau) = 1$ and define a probability matrix $T_n(X,p(\tau))$ by
\ben
\label{eq4.22}
\overline{T}_n(X,p(\tau)) = \sum\limits_{\tau=0}^k p(\tau) \overline{T}_n(X,\tau) \ .
\een
Then the steady state vector of the Markov process with transition matrix $T_n(X,p(\tau))$ is the uniform distribution in which the probability of each class equals the number of states in that class divided by $2^n$, the total number of states in $E_n$.
\end{thm}

\begin{rem}
von Nimwegen, et al.\ \cite{ref36} prove a similar result for the case $p(1) = 1$, $\tau=1$.
\end{rem}

\begin{pf}
Let $m$ be the number of equivalence classes used in the definition of the matrices $\overline{T}_n(X,\tau)$ and let $\vec{v}$ be the $m$-dimensional vector corresponding to a uniform distribution over these classes.  Then
\ben
\label{eq4.23}
\left[ \overline{T}_n(X,\tau)\cdot \vec{v} \right]_i = \sum\limits_{j=1}^m \left[\overline{T}_n(X,\tau)\right]_{ij} v_j \ .
\een

Since $[T_n(X,\tau)]_{ij}$ is the probability that a $\tau$-point mutation from the $j$-th class will be in the $i$-th class, while $v_j$ is the fraction of the total number of states that are contained in the $j$-th class, the sum in (\ref{eq4.23}) is the probability that a state chosen at random from $E_n$ lies in the $i$-th class after a $\tau$-point mutation.  The number of possible mutations of states in $E_n$ is $2^n\binom{n}{\tau}$ while the number of possible mutations from the $i$-th class to $E_n$ is $n_i\tbinom{n}{\tau}$ where $n_i$ is the number of states in this class.  But all mutations are reversible, hence this last number is also the number of $\tau$-point mutations from $E_n$ to the $i$-th class.  Hence the probability that a randomly chosen state of $E_n$ will mutate to a state in the $i$-th class is
\ben
\label{eq4.24}
\frac{n_i\binom{n}{\tau}}{2^n\binom{n}{\tau}} = \frac{n_i}{2^n} = v_i \ .
\een
Thus (\ref{eq4.23}) becomes $\overline{T}_n(X,\tau)\vec{v} = \vec{v}$. \qed
\end{pf}

\section{Relations Between State-Basin and Shift-Basin Representations}
\label{sec5}
Although the shift-basin matrix $\overline{T}_n(X^*,\tau)$ was introduced in Section \ref{sec3}, attention so far has focused on the state-basin transition matrix $\overline{T}_n(X,\tau)$.  In this section the relation between these two matrices is explored.

The equivalence classes $\alpha$:$h(\alpha)$ used to define $\overline{T}_n(X,\tau)$ are sets of states at the same height $h(\alpha)$ above the attractor $\alpha$ in the state transition diagram of the rule $X$.  The classes used in the definition of $\overline{T}_n(X^*,\tau)$ are sets of shift cycles with each shift cycle in the set being at the same height $h(\alpha)$ above the corresponding attractor $\alpha$ in the state-basin diagram.  Since cellular automata rules commute with the shift, the height of a state in the STD is the same height as the shift cycle to which that state belongs in the SBD.  Thus a start at understanding the relation between the matrices $T_n(X,\tau)$ and $\overline{T}_n(X^*,\tau)$ can be found in the relation between the STD and the SBD for the rule $X$.  This relation is most easily explored for additive rules.  In that case, Lemma \ref{lem3.1}, Theorem \ref{thm3.1}, and Lemma \ref{lem3.4} are available as characterizations of the STD, while Theorem \ref{thm3.3} gives a specific connection between this diagram and the shift-basin diagram.

If each of the classes $\alpha$:$h(\alpha)$ in the STD consists of a single shift cycle, then the corresponding transition matrices $\overline{T}_n(X,\tau)$ and $\overline{T}_n(X^*,\tau)$ are equal.  This will be true, for example, for $X = \sigma^k$ for any value of $k$.  In general, however, either at least some of the classes $\alpha$:$h(\alpha)$ will be composed of a union of more than one shift cycle, or will consist of a union of subsets of several shift cycles.  The binary difference rule on a cylinder of size $5$ is an example of the first case --- in addition to the fixed point $\vec{0}$ there is a single attractor which is a period $15$ cycle composed of the union of the three shift cycles $\{\sigma^r(00011)\}$, $\{\sigma^r(00101)\}$, and $\{\sigma^r(01111)\}$ for $0 \le r \le 4$.  The same rule on a cylinder of size $6$ gives an example of the second case: in addition to the fixed point $\vec{0}$ and a period 3 cycle that is a shift cycle, there are two period $6$ cycles the first consisting of the union of the sets $\{\sigma^{2r}(000101)\}$ and $\{\sigma^{2r}(001111)\}$, and the second of the union of the sets $\{\sigma^{2r}(001010)\}$ and $\{\sigma^{2r}(011110)\}$ for $0 \le r \le 2$.

If the classes $\alpha$:$h(\alpha)$ are composed of a union of several full shift cycles then again $\overline{T}_n(X,\tau) = \overline{T}_n(X^*,\tau)$ since the shift classes used in defining $\overline{T}_n(X^*,\tau)$ contain exactly the same states as the corresponding classes $\alpha$:$h(\alpha)$.  Thus, only the case in which the classes $\alpha$:$h(\alpha)$ are composed of the union of proper subsets of shift cycles needs to be considered.  When this is the case, each shift cycle contributes equally if the rule is additive.

\begin{lem}
\label{lem5.1}
Let $X$ be an additive rule defined on a cylinder of size $n$ such that the equivalence classes $\alpha$:$h(\alpha)$ are composed of subsets of two or more shift cycles.  Then, for fixed height $h$, each class $\alpha$:$h(\alpha)$ contains an equal number of elements from each of these shift cycles.
\end{lem}

For this final case, elementary row and column operations can be used to reduce the matrix $\overline{T}_n(X,\tau)$ to the form
\ben
\label{eq5.1}
\overline{T}_n(X,\tau) =
\begin{pmatrix}
\overline{T}_n(X^*,\tau) & 0 \\
B & C
\end{pmatrix} ,
\een
where $C$ is a square matrix.  The characteristic equation for this matrix now becomes
\ben
\label{eq5.2}
\left| \lambda I - C \right| \cdot \left|\lambda I - \overline{T}_n(X^*,\tau)\right| = 0 ,
\een
so that the eigenvalues of $\overline{T}_n(X,\tau)$consist of the eigenvalues of $\overline{T}_n(X^*,\tau)$ and those of the matrix $C$.  In many cases the latter eigenvalues will either be all $0$, or will be the same as some of the eigenvalues of $\overline{T}_n(X^*,\tau)$.

\begin{alg}
\label{alg5.1}
Let $c(\Sigma_i)$ be the set of classes $\alpha_i$:$h(\alpha_i)$, for fixed height $h$, that are composed of states drawn from the shift cycles $S_{ij}$ in the set $\Sigma_i$.
\begin{enumerate}
\item Put $\overline{T}_n(X,\tau)$ into the form of \eqref{eq4.9}
\item The classes $\alpha_i$:$h(\alpha_i)$ label rows and columns of $\overline{T}_n(X,\tau)$.  Let $\alpha_0$:$h(\alpha_0)$ be the label of the first row of $\overline{T}_n(X,\tau)$ corresponding to an element of $c(\Sigma_i)$ and add to this row the remaining rows labelled by members of $c(\Sigma_i)$.  Then move these remaining rows to the bottom of the matrix.
\item Subtract the column labelled by $\alpha_0$:$h(\alpha_0)$ from each of the columns labelled by the remaining elements of $c(\Sigma_i)$, then move these columns to the far right of the matrix.
\item Carry out steps (2) and (3) for each value of the height $h$ and for each set of shift cycles $\Sigma_i$.
\end{enumerate}
\end{alg}

Examples of this algorithm are given in Appendix \ref{ap2}.

\begin{thm}
\label{thm5.1}
Let $X$ be a cellular automata rule defined on a cylinder of size $n$ such that the STD of $X$ contains attractor basins in which there are equivalence classes $\alpha$:$h(\alpha)$ composed of the union of proper subsets of two or more shift cycles.  Then Algorithm \ref{alg5.1} will put the matrix $\overline{T}_n(X,\tau)$ into the form of \eqref{eq5.1}.
\end{thm}

\begin{pf}
All that is required is to show that the operations in step (3) of this algorithm will in fact produce the block of zeros in the matrix of \eqref{eq5.1}.  The remainder of the blocks in this matrix require no explanation: $\overline{T}_n(X^*,\tau)$ arises from the construction set out in the algorithm and the forms of the matrices $B$ and $C$ are not directly specified.

The block of zeros arises when a column labelled $\alpha$:$h(\alpha)$ is subtracted from another column labelled $\beta$:$h(\beta)$ in the case where both equivalence classes so labelled share subsets from the same set of shift cycles.  But this means that if $\mu\in\alpha$:$h(\alpha)$ and $\gamma\in\beta$:$h(\beta)$ are drawn from different subsets of the same shift cycle then there is some fixed $k$ such that $\gamma = \sigma^k(\mu)$.  Thus, both of these states toggle in identical ways, up to a shift, and so will be identical in row positions corresponding to equivalence classes consisting of the union of sets of full shift cycles.  But by construction, these are precisely the classes that label the rows corresponding to the matrix $\overline{T}_n(X^*,\tau)$. This means that the only possible non-zero contributions will appear in the matrix $C$. \qed
\end{pf}

As a result of this theorem we have another immediate result arising from the case in which the matrix $C$ is 0:
\begin{thm}
\label{thm5.2}
Let $X$ be a cellular automata rule on a cylinder of size $n$ such that the matrix $\overline{T}_n(X,\tau)$ has the form of \eqref{eq5.1} with $C = 0$, and let $\tau$ be odd.  Then if $\phi^*(\lambda) = 0$ is the characteristic equation of $\overline{T}_n(X^*,\tau)$, the corresponding characteristic equation of $\overline{T}_n(X,\tau)$ will be $\phi(\lambda) = \lambda^k\phi(\lambda) = 0$ for some $k$.
\end{thm}

In general, if $\phi^*(\lambda) = 0$ and $\phi(\lambda) = 0$ are the characteristic equations of $\overline{T}_n(X^*,\tau)$ and $\overline{T}_n(X,\tau)$ respectively, then $\phi(\lambda) = f(\lambda)\phi^*(\lambda)$.  The conditions under which $f(\lambda)$ has roots that coincide with roots of $\phi^*(\lambda)$ are uncertain.  This is true for the two additive rules described in Appendix \ref{ap2}, but not for the third, non-additive rule described there.  No cases of additive rules for which $T_n(X,\tau)$ and $\overline{T}_n(X,\tau)$  have different eigenvalues have been found.

\begin{conj}
Let $X$ be an additive rule.  Then $\overline{T}_n(X,\tau)$ and $\overline{T}_n(X^*,\tau)$ have the same eigenvalues although the multiplicities of some eigenvalues will differ.
\end{conj}

\section{Discussion}
A number of results have been presented on the use of transition matrices to describe mutational transitions between cellular automata attractor basins, but many questions remain open.  
Appendix \ref{ap1}, for example, lists the characteristic equation for the matrix $\overline{T}_n(X,\tau)$ for a number of additive and non-additive rules.  Inspection shows that the eigenvalues of this matrix for the additive rules considered are simple fractions with denominators related to cycle periods.  This is not the case for the non-additive rules.  Resolution of the questions of whether or not this is generally true is of great interest.  It might be conjectured, for example, that the eigenvalues of $\overline{T}_n(X,\tau)$ for additive rules are always fractions in which the denominator is equal to a cycle period, or to an integer factor of a cycle period.

Another point of interest arises concerning the conjecture at the end of Section \ref{sec5}.  Proof of this conjecture, or determination of the necessary and sufficient conditions for the matrices $\overline{T}_n(X,\tau)$ and $\overline{T}_n(X^*,\tau)$ to have the same eigenvalues would be valuable in understanding the connection between rule state transition diagrams, their shift equivalents, and the toggle relation which is defined on the n-hypercube.  Further analysis of the relation between the structure of a rule STD and the matrix $\overline{T}_n(X,\tau)$ could also shed more light on these connections.

In the case of additive rules, such analysis might prove relatively simple since a $\tau$-point mutation is equivalent to the transformation $\mu\rightarrow \mu+\eta$ where $\eta$ is a state containing $\tau$ ones located at the toggle sites.  For example, the next theorem, on the preservation of vertex structure in state transition diagrams, follows immediately:
\begin{thm}
\label{thm6.1}
Let $X$ be an additive cellular automata rule defined on a cylinder of size $n$.  Let $\mu$ and $\mu'$ be predecessor states of a state $\gamma$, and let $\xi$ and $\xi'$ be $\tau$-point mutations of $\mu$ and $\mu'$: $\xi=\mu+\eta$, $\xi'=\mu'+\eta$.  Then: (1) Both $\xi$ and $\xi'$ are predecessors of the state $\gamma + X(\eta)$; (2) Both $\mu$ and $1+\mu$ mutate to predecessors of the same state $\gamma+X(\eta)$.
\end{thm}

Analysis of special cases can be useful as well.  For example, the binary difference rule has operator form $D = I + \sigma$.  Thus, $D(\mu) = \mu + \sigma(\mu) = D(\vec{1}+\mu)$.  Making use of these relations, it is easy to show that if $\mu$ is at height $h(\mu)$ in the STD of this rule while $D^{h(\mu)-1}(\mu) = \gamma$, then taking $t_{\tau}$ as the operation of making a $\tau$-point mutation,
\ben
\label{eq6.1}
t_{\tau}(\gamma) = t_{\tau}(\mu) + \sigma \sum\limits_{r=0}^{h(\mu)-2} D^r(\mu) \ .
\een

On a more general note, a number of extensions of the work reported in this paper may be possible.  Here the transitions between attractor basins were accomplished by point mutations.  Another possibility is to specify a probability matrix a priori.  This could be used, for example, to model potential wells with probability by allowing transition probabilities depend on the height of a state above the attractor.  Another line of work would be to consider non-cylindrical cellular automata strings of length $mn$ with a random ``heat bath'' at the right end and develop models of fluctuation enhancement processes.  Work along both of these directions is currently in progress.

\begin{ack}
The distinction between fluctuation time scales that are fast or slow with respect to the iteration time scale of a cellular automata rule was brought out during conversations with Andrew Wuensche.  Computations for the matrix for the binary difference rule on a cylinder of size 12 was carried out using Wuensche's Discrete Dynamics Lab.  The hospitality of the University of Arizona Center for Consciousness Studies in December, 2002 and January, 2003; and the hospitality and wonderful research environment provided by the Santa Fe Institute during a four week stay in February and early March, 2003 are gratefully acknowledged.  Conversations with Valerie Watts and David Adams also contributed to the work reported in this paper. Preparations of tables in Appendix \ref{ap1} were assisted by Rhyan Arthur and Carey Luxford, supported by NSERC Undergraduate Student Research Assistantships and the Athabasca University Research Committee. This work was supported by NSERC Discovery Grant OGP 0024817 and by a grant from the Athabasca University Research Committee.
\end{ack}

\addcontentsline{toc}{section}{References}
\bibliographystyle{../../../../base/local/TeX/habbrv}
\bibliography{CA_Transitions}

\appendix

\section{Examples of Characteristic Equations of $\overline{T}_n(X,1)$}
\label{ap1}
Below, characteristic equations are given only for cases in which the size of $\overline{T}_n(X,1)$ is sufficiently small. Also, the third column, $t_M$, gives the maximum period.

\subsection{$\overline{T}_n(D,1)$ for $2\le n\le 13$ (rule $60$)}
\label{ap1.1}
\begin{center}
\begin{tabular}{|c|c|c|}
\hline
$n$ & Characteristic Equation & $t_M$ \\
\hline
 1  & $\lambda^2-1$ & 1 \\
\hline
 2  & $\lambda\left(\lambda^2-1\right)$ & 1 \\
\hline
 3  & $\left(\lambda^2-1\right)\left(\lambda^2-\frac{1}{9}\right)$ & 3 \\
\hline
 4  & $\lambda^3\left(\lambda^2-1\right)$ & 1 \\
\hline
 5  & $\left(\lambda^2-1\right)\left(\lambda^2-\frac{1}{225}\right)$ & 15 \\
\hline
 6  & $\lambda^8\left(\lambda^2-1\right)\left(\lambda^2-\frac{1}{9}\right)$ & 6 \\
\hline
 7  & $\left(\lambda^2-1\right)\left(\lambda^2-\frac{1}{49}\right)^2\left(\lambda^2-\frac{1}{2,401}\right)^7$ & 7 \\
\hline
 8  & $\lambda^7\left(\lambda^2-1\right)$ & 1 \\
\hline
 9  & $\left(\lambda^2-1\right)\left(\lambda^2-\frac{1}{9}\right)\left(\lambda^2-\frac{1}{3,969}\right)\left(\lambda^2-\frac{25}{3,969}\right)^2\left(\lambda^2-\frac{121}{3,969}\right)$ & 63 \\
\hline
 10 & $\lambda^{10}\left(\lambda^2-1\right)\left(\lambda^2-\frac{1}{225}\right)^7\left(\lambda^2-\frac{1}{25}\right)^2$ & 30 \\
\hline
 11 & $\left(\lambda^2-1\right)\left(\lambda^2-\frac{1}{961}\right)^2\left(\lambda^2-\frac{441}{116,281}\right)$ & 341 \\
\hline
 12 & $\lambda^{84}\left(\lambda^2-1\right)\left(\lambda^2-\frac{1}{9}\right)\left(\lambda^2-\frac{1}{16}\right)^4\left(\lambda^2-\frac{1}{144}\right)^{12}$ & 12 \\
\hline
 13 & $\left(\lambda^2-1\right)\left(\lambda^2-\frac{289}{74,529}\right)\left(\lambda^2-\frac{1}{3,969}\right)^4$ & 819 \\
\hline
\end{tabular}
\end{center}

\subsection{$\overline{T}_n(\delta,1)$ for $4\le n$ even $\le 8$ (rule $90$: n odd cases same as rule $60$)}
\label{ap1.2}
\begin{center}
\begin{tabular}{|c|c|c|}
\hline
$n$ & Characteristic Equation & $t_M$ \\
\hline
 4  & $\lambda\left(\lambda-1\right)\left(\lambda+\frac{1}{3}\right)$ & 1 \\
\hline
 6  & $\left(\lambda-1\right)\left(\lambda^2-\frac{1}{9}\right)^2\left(\lambda^2-\frac{1}{81}\right)^2\left(\lambda+\frac{1}{3}\right)^2\left(\lambda-\frac{1}{9}\right)$ & 3 \\
\hline
 8  & $\lambda^3\left(\lambda-1\right)\left(\lambda+\frac{1}{3}\right)$ & 1 \\
\hline
\end{tabular}
\end{center}

\subsection{$\overline{T}_n(\Delta,1)$ for $3\le n\le 11$ (n = $8$, $10$ excluded) (rule $150$)}
\label{ap1.3}
\begin{center}
\begin{tabular}{|c|c|c|}
\hline
$n$ & Characteristic Equation & $t_M$ \\
\hline
 3  & $\left(\lambda^2-1\right)\left(\lambda^2-\frac{1}{9}\right)$ & 1 \\
\hline
 4  & $\lambda^4\left(\lambda^2-1\right)$ & 4 \\
\hline
 5  & $\left(\lambda^2-1\right)\left(\lambda^2-\frac{1}{225}\right)$ & 15 \\
\hline
 6  & $\lambda^5\left(\lambda^2-1\right)\left(\lambda^2-\frac{1}{9}\right)$ & 2 \\
\hline
 7  & $\left(\lambda^2-1\right)\left(\lambda^2-\frac{1}{49}\right)^2\left(\lambda^2-\frac{1}{2,401}\right)^7$ & 7 \\
\hline
 9  & $\left(\lambda^2-1\right)\left(\lambda^2-\frac{1}{9}\right)\left(\lambda^2-\frac{1}{3,969}\right)\left(\lambda^2-\frac{1}{35,721}\right)$ & 63 \\
\hline
11  & $\left(\lambda^2-1\right)\left(\lambda^2-\frac{1}{961}\right)^2\left(\lambda^2-\frac{441}{116,281}\right)$ & 341 \\
\hline
\end{tabular}
\end{center}

\subsection{$\overline{T}_n(18,1)$ for $3\le n\le 7$ (rule $18$)}
\label{ap1.4}
\begin{center}
\begin{tabular}{|c|c|c|}
\hline
$n$ & Characteristic Equation & $t_M$ \\
\hline
 3  & $\left(\lambda-1\right)\left(\lambda^2+\frac{1}{2}\lambda-\frac{1}{6}\right)$ & 1 \\
\hline
 4  & $\left(\lambda-1\right)\left(\lambda^3+\frac{3}{7}\lambda^2-\frac{1}{4}\lambda-\frac{1}{28}\right)$ & 4 \\
\hline
 5  & $\left(\lambda-1\right)\left(\lambda^3+\frac{5}{11}\lambda^2-\frac{279}{1,100}\lambda-\frac{9}{1,100}\right)$ & 10 \\
\hline
 6  & $\left(\lambda-1\right)\left(\lambda^2-\frac{1}{8}\right) \left(\lambda^5+\frac{2\lambda^4}{9}-\frac{131\lambda^3}{324}-\frac{5\lambda^2}{162}+\frac{\lambda}{72}+\frac{1}{5,832}\right)$ & 3 \\
\hline
 7  & $\left(\lambda-1\right)\left(\lambda^5-\frac{457\lambda^4}{1,827}-\frac{1,013\lambda^3}{5,481}+\frac{769\lambda^2}{38,367}+\frac{1,712\lambda}{268,569}-\frac{128}{626,661}\right)$ & 1 \\
\hline
\end{tabular}
\end{center}

\subsection{$\overline{T}_n(22,1)$ for $3\le n\le 6$ (rule $22$)}
\label{ap1.5}
\begin{center}
\begin{tabular}{|c|c|c|}
\hline
$n$ & Characteristic Equation & $t_M$ \\
\hline
 3  & $\left(\lambda-1\right)\left(\lambda^2+\frac{1}{2}\lambda-\frac{1}{6}\right)$ & 1 \\
\hline
 4  & $\lambda\left(\lambda-1\right)\left(\lambda^3+\frac{3}{5}\lambda^2-\frac{13}{40}\lambda-\frac{3}{40}\right)$ & 4 \\
\hline
 5  & $\left(\lambda-1\right)\left(\lambda^2-\frac{1}{25}\right)\left(\lambda^4+\frac{2\lambda^3}{3}-\frac{12\lambda^2}{25}-\frac{14\lambda}{75}+\frac{3}{125}\right)$ & 1 \\
\hline
 6  & $\left(\lambda-1\right)\left(\lambda^6+\frac{\lambda^5}{10}-\frac{117\lambda^4}{324}+\frac{59\lambda^3}{6,480}+\frac{1,009\lambda^2}{38,880}-\frac{7\lambda}{2,592}-\frac{1}{19,440}\right)$ & 2 \\
\hline
\end{tabular}
\end{center}

\subsection{$\overline{T}_n(30,1)$ for $3\le n\le 7$ (rule $30$)}
\label{ap1.6}
\begin{center}
\begin{tabular}{|c|c|c|}
\hline
$n$ & Characteristic Equation & $t_M$ \\
\hline
 3  & $\left(\lambda^2-1\right)\left(\lambda^2-\frac{1}{9}\right)$ & 1 \\
\hline
 4  & $\lambda^3\left(\lambda^2-1\right)$ & 8 \\
\hline
 5  & $\left(\lambda^2-1\right)\left(\lambda^2-\frac{9}{25}\right)\left(\lambda^2-\frac{1}{25}\right)^2$ & 5 \\
\hline
\raisebox{-2ex}[0pt]{$6$} & \multicolumn{1}{|l|}{$\lambda^2\left(\lambda-1\right)\left(\lambda^2-\frac{1}{9}\right)\cdot$} & \raisebox{-2ex}[0pt]{$2$} \\
& \multicolumn{1}{|r|}{$\left(\lambda^7+\lambda^6-\frac{7\lambda^5}{48}-\frac{117\lambda^4}{432}-\frac{7\lambda^3}{1,296}+\frac{7\lambda^2}{432}+\frac{\lambda}{3,888}-\frac{1}{3,888}\right)$} & \\
\hline
 7  & $\lambda^2\left(\lambda-1\right)\left(\lambda^2-\frac{1}{9}\right)\left(\lambda^5+\frac{31\lambda^4}{63}-\frac{103\lambda^3}{882}-\frac{23\lambda^2}{686}+\frac{3\lambda}{2,744}+\frac{229}{1,210,104}\right)$ & 63 \\
\hline
\end{tabular}
\end{center}

\subsection{$\overline{T}_n(54,1)$ for $3\le n\le 6$ (rule $54$)}
\label{ap1.7}
\begin{center}
\begin{tabular}{|c|c|c|}
\hline
$n$ & Characteristic Equation & $t_M$ \\
\hline
 3  & $\left(\lambda^2-1\right)\left(\lambda^2-\frac{1}{9}\right)$ & 1 \\
\hline
 4  & $\lambda^6\left(\lambda^2-1\right)$ & 4 \\
\hline
 5  & $\lambda\left(\lambda^2-1\right)\left(\lambda^2-\frac{1}{25}\right)\left(\lambda^2-\frac{1}{5}\right)$ & 1 \\
\hline
 6  & $\lambda^2\left(\lambda-1\right)\left(\lambda^2-\frac{\lambda}{4}-\frac{1}{16}\right) \left(\lambda^5+\frac{3\lambda^4}{4}-\frac{7\lambda^3}{240}-\frac{341\lambda^2}{6,480}-\frac{53\lambda}{38,880}+\frac{5}{10,976}\right)$ & 12 \\
\hline
\end{tabular}
\end{center}

\subsection{$\overline{T}_n(110,1)$ for $3\le n\le 6$ (rule $110$)}
\label{ap1.8}
\begin{center}
\begin{tabular}{|c|c|c|}
\hline
$n$ & Characteristic Equation & $t_M$ \\
\hline
 3  & $\left(\lambda^2-1\right)\left(\lambda^2-\frac{1}{9}\right)$ & 1 \\
\hline
 4  & $\lambda^6\left(\lambda^2-1\right)\left(\lambda^2-\frac{1}{4}\right)$ & 4 \\
\hline
 5  & $\lambda^2\left(\lambda-1\right)\left(\lambda^3+\frac{3\lambda^2}{5}-\frac{6\lambda}{25}-\frac{2}{25}\right)$ & 1 \\
\hline
 6  & $\left(\lambda-1\right)\left(\lambda^7+\frac{7\lambda^6}{9}+\frac{67\lambda^5}{1,620}-\frac{7\lambda^4}{120}-\frac{19\lambda^3}{4,320}+\frac{61\lambda^2}{69,984}+\frac{7\lambda}{349,920}-\frac{1}{349,920}\right)$ & 18 \\
\hline
\end{tabular}
\end{center}

\section{Examples of Algorithm \ref{alg5.1} for $\tau = 1$}
\label{ap2}

\begin{exmp}[Binary difference rule on a cylinder of size 6]
\label{exmp1}
Let $D$ be the global operator for the binary difference rule. On a cylinder of size 6 the equivalence classes of states are:
\begin{center}
\begin{tabular}{|l|}
\hline
$0$:$0$ = $\{000000\}$ \\
$0$:$1$ = $\{111111\}$ \\
$0$:$2$ = $\{010101$, $101010\}$ \\
\hline
$1$:$0$ = $\{\sigma^r(000101)$, $\sigma^r(001111)|0\le r\le 2\}$ \\
$1$:$1$ = $\{\sigma^r(000011)$, $\sigma^r(111010)|0\le r\le 2\}$ \\
$1$:$2$ = $\{\sigma^r(000001)$, $\sigma^r(111110)|0\le r\le 2\}$ \\
\hline
$2$:$0$ = $\{\sigma^r(001010)$, $\sigma^r(011110)|0\le r\le 2\}$ \\
$2$:$1$ = $\{\sigma^r(000110)$, $\sigma^r(110101)|0\le r\le 2\}$ \\
$2$:$2$ = $\{\sigma^r(000010)$, $\sigma^r(111101)|0\le r\le 2\}$ \\
\hline
$3$:$0$ = $\{\sigma^r(011011)|0\le r\le 2\}$ \\
$3$:$1$ = $\{\sigma^r(001001)|0\le r\le 2\}$ \\
$3$:$2$ = $\{\sigma^r(111000)|0\le r\le 2\}$ \\
\hline
\end{tabular}
\end{center}

The equivalence classes sharing states from the same shift cycles are $(1$:$0$,$2$:$0)$, $(1$:$1$,$2$:$1)$, and $(1$:$2$,$2$:$2)$.  In the form of \eqref{eq4.9} the matrix $\overline{T}_n(D,1)$ is
\be
\left(\begin{array}{cccccccccccc}
\multicolumn{8}{c}{} & 0 & \frac{1}{24} & \frac{1}{24} &  0 \\
\multicolumn{8}{c}{} & 0 & \frac{1}{24} & \frac{1}{24} &  0 \\
\multicolumn{8}{c}{} & \frac{1}{4} & \frac{1}{6} & \frac{1}{6} & \frac{1}{4} \\
\multicolumn{8}{c}{\raisebox{-1.5ex}[0pt]{$0$}} & \frac{1}{4} & \frac{1}{6} & \frac{1}{6} & \frac{1}{4} \\
\multicolumn{8}{c}{} & \frac{1}{4} & \frac{1}{6} & \frac{1}{6} & \frac{1}{4} \\
\multicolumn{8}{c}{} & \frac{1}{4} & \frac{1}{6} & \frac{1}{6} & \frac{1}{4} \\
\multicolumn{8}{c}{} & 0 & \frac{1}{8} & \frac{1}{8} &  0 \\
\multicolumn{8}{c}{} & 0 & \frac{1}{8} & \frac{1}{8} &  0 \\
0 & 0 & \frac{1}{12} & \frac{1}{12} & \frac{1}{12} & \frac{1}{12} & 0 & 0 & \multicolumn{4}{c}{} \\
\frac{1}{2} & \frac{1}{2} & \frac{1}{3} & \frac{1}{3} & \frac{1}{3} & \frac{1}{3} & \frac{1}{2} & \frac{1}{2} & \multicolumn{4}{c}{\raisebox{-1.5ex}[0pt]{$0$}} \\
\frac{1}{2} & \frac{1}{2} & \frac{1}{3} & \frac{1}{3} & \frac{1}{3} & \frac{1}{3} & \frac{1}{2} & \frac{1}{2} & \multicolumn{4}{c}{} \\
0 & 0 & \frac{1}{4} & \frac{1}{4} & \frac{1}{4} & \frac{1}{4} & 0 & 0 & \multicolumn{4}{c}{}
\end{array}\right) \ .
\ee

The indices for both rows and columns are ordered $0$:$0$, $0$:$1$, $1$:$0$, $1$:$1$, $2$:$0$, $2$:$1$, $3$:$0$, $3$:$1$, $0$:$2$, $1$:$2$, $2$:$2$, $3$:$2$. Add the row labelled $2$:$0$ to that labelled $1$:$0$, the row labelled $2$:$1$ to that labelled $1$:$1$, and the row labelled $2$:$2$ to that labelled $1$:$2$. Then move rows $2$:$0$, $2$:$1$, and $2$:$2$ to the bottom of the matrix. This yields
\be
\left(\begin{array}{cccccccccccc}
\multicolumn{8}{c}{} & 0   & \frac{1}{24} & \frac{1}{24} &  0   \\
\multicolumn{8}{c}{} & 0   & \frac{1}{24} & \frac{1}{24} &  0   \\
\multicolumn{8}{c}{\raisebox{-1.5ex}[0pt]{$0$}} & \frac{1}{2} & \frac{1}{3} & \frac{1}{3} & \frac{1}{2} \\
\multicolumn{8}{c}{} & \frac{1}{2} & \frac{1}{3} & \frac{1}{3} & \frac{1}{2} \\
\multicolumn{8}{c}{} & 0   & \frac{1}{8} & \frac{1}{8} &  0   \\
\multicolumn{8}{c}{} & 0   & \frac{1}{8} & \frac{1}{8} &  0   \\
0 & 0 & \frac{1}{12} & \frac{1}{12} & \frac{1}{12} & \frac{1}{12} & 0 & 0 & \multicolumn{4}{c}{} \\
1 & 1 & \frac{2}{3} & \frac{2}{3} & \frac{2}{3} & \frac{2}{3} & 1 & 1 & \multicolumn{4}{c}{0} \\
0 & 0 & \frac{1}{4} & \frac{1}{4} & \frac{1}{4} & \frac{1}{4} & 0 & 0 & \multicolumn{4}{c}{} \\
\multicolumn{8}{c}{\raisebox{-1.5ex}[0pt]{$0$}} & \frac{1}{4} & \frac{1}{6} & \frac{1}{6} & \frac{1}{4} \\
\multicolumn{8}{c}{} & \frac{1}{4} & \frac{1}{6} & \frac{1}{6} & \frac{1}{4} \\
\frac{1}{2} & \frac{1}{2} & \frac{1}{3} & \frac{1}{3} & \frac{1}{3} & \frac{1}{3} & \frac{1}{2} & \frac{1}{2} & \multicolumn{4}{c}{0} \\
\end{array}\right) \ .
\ee

Now subtract the column labelled $1$:$0$ from column $2$:$0$, the column labelled $1$:$1$ from column $2$:$1$, and the column labelled $1$:$2$ from column $2$:$2$. Following this, shift columns $2$:$0$, $2$:$1$, and $2$:$2$ to the far right of the matrix. The result is
\be
\left(\begin{array}{cccccccccccc}
\multicolumn{6}{c}{} & 0 & \frac{1}{24} & 0 & 0 & 0 & 0 \\
\multicolumn{6}{c}{} & 0 & \frac{1}{24} & 0 & 0 & 0 & 0 \\
\multicolumn{6}{c}{\raisebox{-1.5ex}[0pt]{$0$}} & \frac{1}{2} & \frac{1}{3} & \frac{1}{2} & 0 & 0 & 0 \\
\multicolumn{6}{c}{} & \frac{1}{2} & \frac{1}{3} & \frac{1}{2} & 0 & 0 & 0 \\
\multicolumn{6}{c}{} & 0 & \frac{1}{8} & 0 & 0 & 0 & 0 \\
\multicolumn{6}{c}{} & 0 & \frac{1}{8} & 0 & 0 & 0 & 0 \\
0 & 0 & \frac{1}{12} & \frac{1}{12} & 0 & 0 & 0 & 0 & 0 & 0 & 0 & 0 \\
1 & 1 & \frac{2}{3} & \frac{2}{3} & 1 & 1 & 0 & 0 & 0 & 0 & 0 & 0 \\
0 & 0 & \frac{1}{4} & \frac{1}{4} & 0 & 0 & 0 & 0 & 0 & 0 & 0 & 0 \\
0 & 0 & 0 & 0 & 0 & 0 & \frac{1}{4} & \frac{1}{6} & \frac{1}{4} & 0 & 0 & 0 \\
0 & 0 & 0 & 0 & 0 & 0 & \frac{1}{4} & \frac{1}{6} & \frac{1}{4} & 0 & 0 & 0 \\
\frac{1}{2} & \frac{1}{2} & \frac{1}{3} & \frac{1}{3} & \frac{1}{2} & \frac{1}{2} & 0 & 0 & 0 & 0 & 0 & 0
\end{array}\right) ,
\ee
which is in the form of \eqref{eq5.1} with the upper $9\times9$ block the matrix $\overline{T}_n(X^*,1)$.
\end{exmp}

\begin{exmp}[Rule 90 on a cylinder of size 6]
\label{exmp2}
Let $\delta$ represent the global operator for rule 90.  On a cylinder of size 6 the equivalence classes for $\overline{T}_n(\delta,1)$ are
\begin{center}
\begin{tabular}{|lcl|}
\hline
$0$:$0$ &=& $\{000000\}$ \\
$0$:$1$ &=& $\{111111$, $010101$, $101010\}$ \\
\hline
$1$:$0$ &=& $\{\sigma^r(011011)| 0 \le r \le 2\}$ \\
$1$:$1$ &=& $\{\sigma^r(001001)|0\le r\le2\}\cup\{\sigma^r(000111)|0\le r\le5\}$ \\
\hline
$2$:$0$ &=& $\{000101$, $010001$, $010100\}$ \\
$2$:$1$ &=& $\{\sigma^2(000001)|0\le r\le2\}\cup\{\sigma^2(111110)|0\le r\le2\}$ \\
        & & $\cup\{\sigma^2(101011)|0\le r\le2\}$ \\
\hline
$3$:$0$ &=& $\{001010$, $100010$, $101000\}$ \\
$3$:$1$ &=& $\{\sigma^2(000010)|0\le r\le2\}\cup\{\sigma^2(111101)|0\le r\le2\}$ \\
        & & $\cup\{\sigma^2(010111)|0\le r\le2\}$ \\
\hline
$4$:$0$ &=& $\{111100$, $001111$, $110011\}$ \\
$4$:$1$ &=& $\{\sigma^2(000011)|0\le r\le2\}\cup\{\sigma^2(100110)|0\le r\le2\}$ \\
        & & $\cup\{\sigma^2(011001)|0\le r\le2\}$ \\
\hline
$5$:$0$ &=& $\{111001$, $100111$, $011110\}$ \\
$5$:$1$ &=& $\{\sigma^2(000110)|0\le r\le2\}\cup\{\sigma^2(001101)|0\le r\le2\}$ \\
        & & $\cup\{\sigma^2(110010)|0\le r\le2\}$ \\
\hline
\end{tabular}
\end{center}
Classes sharing states from the same shift cycles are: ($2$:$0$,$3$:$0$), ($2$:$1$,$3$:$1$), ($4$:$0$,$5$:$0$), and ($4$:$1$,$5$:$1$).  The matrix $\overline{T}_n(\delta,1)$ with interior/peripheral indexing in the order $0$:$0$, $1$:$0$, $2$:$0$, $3$:$0$, $4$:$0$, $5$:$0$, $0$:$1$, $1$:$1$, $2$:$1$, $3$:$1$, $4$:$1$, $5$:$1$ is
\be
\left(\begin{array}{cccccccccccc}
\multicolumn{6}{c}{} & 0 & 0 & \frac{1}{18} & \frac{1}{18} & 0 & 0 \\
\multicolumn{6}{c}{} & 0 & 0 & \frac{1}{18} & \frac{1}{18} & \frac{1}{9} & \frac{1}{9} \\
\multicolumn{6}{c}{\raisebox{-1.5ex}[0pt]{$0$}} & \frac{1}{6} & \frac{1}{18} & \frac{1}{9} & 0 & \frac{1}{18} & \frac{1}{18} \\
\multicolumn{6}{c}{} & \frac{1}{6} & \frac{1}{18} & 0 & \frac{1}{9} & \frac{1}{18} & \frac{1}{18} \\
\multicolumn{6}{c}{} & 0 & \frac{1}{9} & \frac{1}{18} & \frac{1}{18} & 0 & \frac{1}{9} \\
\multicolumn{6}{c}{} & 0 & \frac{1}{9} & \frac{1}{18} & \frac{1}{18} & \frac{1}{9} & 0 \\
0 & 0 & \frac{1}{6} & \frac{1}{6} & 0 & 0 & 0 & 0 & \frac{1}{9} & \frac{1}{9} & 0 & 0 \\
0 & 0 & \frac{1}{6} & \frac{1}{6} & \frac{1}{3} & \frac{1}{3} & 0 & 0 & \frac{1}{9} & \frac{1}{9} & \frac{2}{9} & \frac{2}{9} \\
\frac{1}{2} & \frac{1}{6} & \frac{1}{3} & 0 & \frac{1}{6} & \frac{1}{6} & \frac{1}{3} & \frac{1}{9} & \frac{2}{9} & 0 & \frac{1}{9} & \frac{1}{9} \\
0 & \frac{1}{3} & \frac{1}{6} & \frac{1}{6} & 0 & \frac{1}{3} & 0 & \frac{2}{9} & \frac{1}{9} & \frac{1}{9} & 0 & \frac{2}{9} \\
0 & \frac{1}{3} & \frac{1}{6} & \frac{1}{6} & \frac{1}{3} & 0 & 0 & \frac{2}{9} & \frac{1}{9} & \frac{1}{9} & \frac{2}{9} & 0
\end{array}\right) .
\ee

Carrying out the operations indicated: adding the row labelled by $3$:$0$ to that labelled by $2$:$0$; by $3$:$1$ to that labelled by $2$:$1$;  by $5$:$0$ to that labelled by $4$:$0$; and by $5$:$1$ to that labelled by $4$:$1$ then moving rows $3$:$0$, $3$:$1$, $5$:$0$, and $5$:$1$ to the bottom of the matrix; then subtracting column $2$:$0$ from $3$:$0$, $2$:$1$ from $3$:$1$, $4$:$0$ from $5$:$0$, and $4$:$1$ from $5$:$1$ followed by moving columns $3$:$0$, $3$:$1$, $5$:$0$, and $5$:$1$ to the far right of the matrix yields the form
\be
\left(\begin{array}{cccccccccccc}
\multicolumn{4}{c}{} & 0 & 0 & \frac{1}{18} & 0 & \multicolumn{4}{c}{} \\
\multicolumn{4}{c}{\raisebox{-1.5ex}[0pt]{$0$}} & 0 & 0 & \frac{1}{18} & \frac{1}{9} & \multicolumn{4}{c}{\raisebox{-1.5ex}[0pt]{$0$}} \\
\multicolumn{4}{c}{} & \frac{1}{3} & \frac{1}{9} & \frac{1}{9} & \frac{1}{9} & \multicolumn{4}{c}{} \\
\multicolumn{4}{c}{} & 0 & \frac{2}{9} & \frac{1}{9} & \frac{1}{9} & \multicolumn{4}{c}{} \\
0 & 0 & \frac{1}{6} & 0 & 0 & 0 & \frac{1}{9} & 0 & \multicolumn{4}{c}{} \\
0 & 0 & \frac{1}{6} & \frac{1}{3} & 0 & 0 & \frac{1}{9} & \frac{2}{9} & \multicolumn{4}{c}{\raisebox{-1.5ex}[0pt]{$0$}} \\
1 & \frac{1}{3} & \frac{1}{3} & \frac{1}{3} & \frac{2}{3} & \frac{2}{9} & \frac{2}{9} & \frac{2}{9} & \multicolumn{4}{c}{} \\
0 & \frac{2}{3} & \frac{1}{3} & \frac{1}{3} & 0 & \frac{4}{9} & \frac{2}{9} & \frac{2}{9} & \multicolumn{4}{c}{} \\
0 & 0 & 0 & 0 & \frac{1}{6} & \frac{1}{18} & 0 & \frac{1}{18} & 0 & 0 & \frac{1}{9} & 0 \\
0 & 0 & 0 & 0 & 0 & \frac{1}{9} & \frac{1}{18} & \frac{1}{9} & 0 & 0 & 0 & -\frac{1}{9} \\
\frac{1}{2} & \frac{1}{6} & 0 & \frac{1}{6} & \frac{1}{3} & \frac{1}{9} & 0 & \frac{1}{9} & \frac{1}{3} & 0 & \frac{2}{9} & 0 \\
0 & \frac{1}{3} & \frac{1}{6} & \frac{1}{3} & 0 & \frac{2}{9} & \frac{1}{9} & \frac{2}{9} & 0 & -\frac{1}{3} & 0 & -\frac{2}{9} \\
\end{array}\right) \ .
\ee

Again this has the form of \eqref{eq5.1}.  For the binary difference rule
\be 
\left|\lambda I-\overline{T}_6(\delta,1)\right| = \lambda^3\left|\lambda I-\overline{T}_6(\delta^*,1)\right| \ .
\ee
For rule $90$ the matrix $C$ is not zero and
\be
\left|\lambda I-\overline{T}_6(\delta,1)\right| = \left(\lambda^2 - 1\right) \left(\lambda^2 - \frac{1}{81} \right) \left|\lambda I-\overline{T}_6(\delta^*,1)\right| \ .
\ee
In both of these cases, computation of eigenvalues shows that the state-basin and shift-basin matrices have the same eigenvalues.  The next example demonstrates that this is not always the case.
\end{exmp}

\begin{exmp}[Rule 54 on a Cylinder of Size 6]
\label{exmp3}
Figure \ref{figA2.1} shows the state transition diagram for rule 54 on a cylinder of size 6.
\begin{figure}
\begin{center}
\end{center}
\resizebox{\linewidth}{!}{\includegraphics{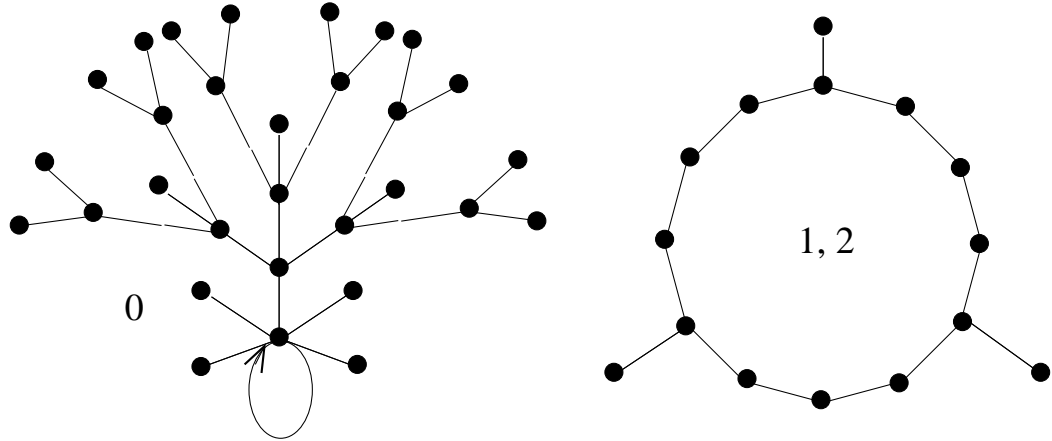}}
\caption{State transition diagram for left justified rule $54$ on a cylinder of size $6$. The numbers accompanying each diagram enumerate the cycle structures. The presence of more than one number indicates that several cycles have the same structure.}
\label{figA2.1}
\end{figure}

The equivalence classes drawn from this STD are:
\begin{center}
\begin{tabular}{|lcl|}
\hline
$0$:$0$ &=& $\{000000\}$ \\
$0$:$1$ &=& $\{111111$, $010101$, $101010\}$ \\
$0$:$2$ &=& $\{001001$, $010010$, $100100\}$ \\
$0$:$3$ &=& $\{\sigma^r(000011)| 0 \le r \le 5\}$ \\
$0$:$4$ &=& $\{\sigma^r(001111)| 0 \le r \le 5\}$ \\
$0$:$5$ &=& $\{\sigma^r(101100)\}\cup\{\sigma^r(100110)\}\ 0 \le r \le 5$ \\
\hline
$1$:$0$ &=& $\{\sigma^{2r}(000001)\}\cup\{\sigma^{2r}(000111)\}\cup\{\sigma^{2r}(010001)\}\cup\{\sigma^{2r}(011111)\}$ \\
        & & $\ \ 0 \le r \le 2$ \\
$1$:$1$ &=& $\{\sigma^{2r}(011101)\}\ 0 \le r \le 2$ \\
\hline
$2$:$0$ &=& $\{\sigma^{2r}(000010)\}\cup\{\sigma^{2r}(001110)\}\cup\{\sigma^{2r}(100010)\}\cup\{\sigma^{2r}(111110)\}$ \\
        & & $\ \ 0 \le r \le 2$ \\
$2$:$1$ &=& $\{\sigma^{2r}(111010)\}\ 0 \le r \le 2$ \\
\hline
\end{tabular}
\end{center}

Taking the row and column indexing in the order: $0$:$0$, $0$:$1$, $0$:$3$, $0$:$4$, $1$:$1$, $2$:$1$, $0$:$2$, $0$:$5$, $1$:$0$, $2$:$0$, the matrix $\overline{T}_n(54,1)$ has the form
\be
\left(\begin{array}{cccccccccc}
\multicolumn{6}{c}{} & 0 & 0 & \frac{1}{24} & \frac{1}{24} \\
\multicolumn{6}{c}{\raisebox{-1.5ex}[0pt]{$0$}} & 0 & 0 & \frac{1}{24} & \frac{1}{24} \\
\multicolumn{6}{c}{} & 0 & \frac{1}{3} & \frac{5}{24} & \frac{5}{24} \\
\multicolumn{6}{c}{} & 0 & \frac{1}{6} & \frac{1}{6} & \frac{1}{6} \\
0 & 0 & 0 & 0 & \frac{1}{6} & \frac{1}{6} & 0 & \frac{1}{6} & \frac{1}{2} & \frac{1}{2} \\
0 & 0 & \frac{4}{9} & \frac{1}{3} & \frac{1}{3} & \frac{1}{3} & \frac{2}{5} & 0 & \frac{1}{12} & \frac{1}{12} \\
\frac{1}{2} & \frac{1}{2} & \frac{5}{18} & \frac{1}{3} & \frac{1}{2} & 0 & \frac{1}{5} & \frac{1}{12} & \frac{1}{4} & 0 \\
\frac{1}{2} & \frac{1}{2} & \frac{5}{18} & \frac{1}{3} & 0 & \frac{1}{2} & \frac{1}{5} & \frac{1}{12} & 0 & \frac{1}{4}
\end{array}\right) \ .
\ee

The equivalence classes sharing shift cycles are ($1$:$0$, $2$:$0$) and ($1$:$1$, $2$:$1$).  Application of Algorithm \ref{alg5.1} yields the matrix
\be
\left(\begin{array}{cccccccccc}
\multicolumn{5}{c}{} & 0 & 0 & \frac{1}{24} & \multicolumn{2}{c}{} \\
\multicolumn{5}{c}{} & 0 & 0 & \frac{1}{24} & \multicolumn{2}{c}{} \\
\multicolumn{5}{c}{0} & 0 & \frac{1}{3} & \frac{5}{24} & \multicolumn{2}{c}{0} \\
\multicolumn{5}{c}{} & 0 & \frac{1}{6} & \frac{1}{6} & \multicolumn{2}{c}{} \\
\multicolumn{5}{c}{} & \frac{1}{5} & \frac{1}{6} & \frac{1}{8} & \multicolumn{2}{c}{} \\
0 & 0 & 0 & 0 & \frac{1}{6} & 0 & \frac{1}{6} & \frac{1}{12} & \multicolumn{2}{c}{} \\
0 & 0 & \frac{4}{9} & \frac{1}{3} & \frac{1}{3} & \frac{2}{5} & 0 & \frac{1}{12} & \multicolumn{2}{c}{0} \\
1 & 1 & \frac{5}{9} & \frac{2}{3} & \frac{1}{2} & \frac{2}{5} & \frac{1}{6} & \frac{1}{4} & \multicolumn{2}{c}{} \\
0 & 0 & 0 & 0 & 0 & \frac{1}{10} & \frac{1}{12} & 0 & 0 & \frac{1}{8} \\
\frac{1}{2} & \frac{1}{2} & \frac{5}{18} & \frac{1}{3} & 0 & \frac{1}{5} & \frac{1}{12} & 0 & \frac{1}{2} & \frac{1}{4}
\end{array}\right) \ .
\ee

The $8\times8$ matrix in the upper left is the matrix $\overline{T}_n(54^*,1)$ having the characteristic equation
\be
\phi^*(\lambda) = \lambda^2 \left(\lambda - 1\right) \left(\lambda^5 + \frac{3}{4}\lambda^4 - \frac{7}{240}\lambda^3 - \frac{341}{6,480}\lambda^2 - \frac{53}{38,880}\lambda + \frac{25}{38,880} \right) .
\ee
On the other hand, the characteristic equation for $\overline{T}_n(54,1)$ is
\be
\phi(\lambda) = \left( \lambda^2 - \frac{1}{4}\lambda - \frac{1}{16} \right) \phi^*(\lambda) .
\ee
The roots of the quadratic factor are $\left(1\pm\sqrt{5}\right)/8 (.404508, -.154508)$ while the roots of $\phi^*(\lambda)$ are $.6883559$, $.2331465$, $.1031093$, $-.1649678$, and $-.2337321$.  This shows that there are rules for which the roots of $\phi(\lambda)$ and $\phi^*(\lambda)$ are not the same.  In the first two cases, the rules were additive, while rule 54 is not additive.
\end{exmp}

\end{document}